\documentclass[%
 reprint,
 amsmath,amssymb,
 aps,
]{revtex4-2}

\usepackage{graphicx}
\usepackage{dcolumn}
\usepackage{bm}

\usepackage[final]{microtype}

\usepackage{amsmath, amssymb, amscd, amsthm, amsfonts,utfsym}
\usepackage{mathrsfs}
\usepackage[cal=cm,scr=boondoxo]{mathalfa} 
\usepackage{graphicx}
\usepackage{hyperref}
\usepackage{xcolor}
\usepackage{cleveref}
\usepackage{parskip}
\usepackage{tikz}
\usepackage{csquotes} 
\usepackage{makecell}
\usepackage{graphicx}
\usepackage{bigstrut}
\usepackage{multirow}
\usepackage{array}

\usepackage{csquotes} 
\usepackage{makecell}
\usepackage{graphicx}
\usepackage{bigstrut}
\usepackage{multirow}
\usepackage{braket}

\usepackage[english]{babel}
\makeatletter
\providecommand{\l@eng}{\l@english}
\makeatother

\usepackage[T1]{fontenc}
\usepackage[utf8]{inputenc} 
\usepackage{lmodern}

\newtheorem{theorem}{Theorem}[section]
\newtheorem{lemma}[theorem]{Lemma}
\newtheorem{corollary}[theorem]{Corollary}
\newtheorem{proposition}[theorem]{Proposition}

\theoremstyle{definition}
\newtheorem{remark}[theorem]{Remark}

\newcommand{\rr}{\mathbb{R}}
\newcommand{\cc}{\mathbb{C}}

\newcommand{\al}{\alpha}

\usepackage{mathrsfs}
\usepackage{array}  

\usepackage[final]{microtype}
\newcommand{\mix}[2]{%
  \makebox[1cm][r]{$#1$}\,$-$\,\makebox[1.8cm][l]{$#2$}%
}
\newcommand{\mixhead}{%
  \makebox[2cm][c]{Mixture}%
}

\newcommand{\ve}{\varepsilon}
\newcommand{\hh}{\mathcal H}
\newcommand{\sss}{\mathcal S}
\newcommand{\dd}{\mathrm{d}}
\newcommand{\ee}{\mathrm{e}}

\newcommand{\pd}[2]{\frac{\partial {#1}}{\partial {#2}}}

\newcommand{\R}{\bf R}

\newcommand{\x}{{\bf  x}}
\newcommand{\kk}{{\bf k}}

\newcommand{\y}{{\bf  y}}

\renewcommand{\Im}{\operatorname{Im}\,}

\begin{document}
\preprint{APS/123-QED}

\title{Efimov spectrum in the Born--Oppenheimer picture of $2+1$ system \\ with zero-range heavy-light interactions}

\author{Hamidreza Saberbaghi}
 \email{hamidreza.saberbaghi@uninsubria.it}
\affiliation{University of Insubria, Science and High Technology Department, Via Valleggio 11, Como, 22100, , Italy
}%





\begin{abstract}
We study the Born--Oppenheimer approximation of a mass-imbalanced three-body system made of two heavy particles of mass $M$ and one light particle of mass $m$ for arbitrary angular momentum. In this system, heavy–light pairs interact via a zero-range force. We construct the light-particle Hamiltonian using self-adjoint extensions of the two-center point interaction and show that the corresponding effective potential is regular at the coincidence point of the heavy particles. Consequently, this model presents an
alternative method to finite-range, cutoff, or short-distance heavy-heavy regularizations: the necessary three-body input is encoded in the self-adjoint realization of the light-particle Hamiltonian, while the heavy-light interactions remain point-like.

In the unitary limit, after fixing the characteristic length scale, we derive an explicit Efimov spectrum. Our results recover the zero-angular-momentum case of \cite{FST} and provide a sufficient condition ensuring the absence of non-Efimov bound states.
 
Away from unitarity, we show that the spatial size of the shallowest trimer near the threshold is approximately 
$2.8$ times the heavy–light scattering length, in contrast to the common assumption that these two length scales coincide. We also derive a Bargmann-type bound on the number of three-body bound states and obtain an estimate sharper than previous results. Finally, we illustrate the method with numerical results for selected alkali mixtures.
\end{abstract}

\maketitle

\section{Introduction}
The Efimov effect \cite{Efimov1,Efimov2} is a universal low-energy feature of three-body systems in $\mathbb{R}^3$: when at least two of the three pairs are at resonance, \emph{i.e.}, in the unitary limit with infinite two-body $S$-wave scattering length $a$, the three-body Hamiltonian supports an infinite sequence of shallow trimers $E_n$ with relatively large spatial sizes, accumulating at zero energy, even though the individual two-body subsystems are not bound. Asymptotically, the ratio of successive binding energies approaches a universal geometric factor,
\begin{equation}\label{universal_geometrical_law}
\lim_{n \rightarrow \infty} \frac{E_n}{E_{n+1}} = \ee^{\tfrac{2 \pi}{s_0}} \, ,
\end{equation}
where $s_0$ is a dimensionless parameter that depends on the number of resonant pairs, the mass ratio of the particles, and their statistics.
 
Experimentally, weakly bound three-body states with large spatial extent have been observed in mass-imbalanced mixtures such as $^{133}\mathrm{Cs}_2-^{6}\mathrm{Li}$ \cite{PhysRevLett.113.240402,PhysRevLett.112.250404} and in systems of identical bosons (see \cite{kolganova20174} and the references therein). In ultracold atomic gases, the large two-body scattering lengths required to access this regime are realized using magnetic Feshbach resonances \cite{RevModPhys.82.1225}, which have also been exploited recently in mass-imbalanced 
$^{167}\mathrm{Er}_2-^{6}\mathrm{Li}$,  $^{166}\mathrm{Er}_2-^{6}\mathrm{Li}$, and
$^{164}\mathrm{Er}_2-^{6}\mathrm{Li}$ mixtures in the microkelvin regime \cite{doi:10.7566/JPSJ.92.054301,kalia2025creation,Xie2025LiDyFeshbach}.
 
Theoretically, these observations are often described within zero-range models. At low energies, scattering properties are largely independent of the microscopic details of the short-range forces \cite{BRAATEN2006259}, so zero-range potentials (Fermi pseudo-potentials \cite{FermiNeutron,PhysRev.71.215}) provide a natural framework. The universal behavior of such systems is characterized by the two-body sub-systems and their $S$-wave scattering length, determined as the zero-energy limit of the scattering amplitude. A standard example of this universality obtained in zero-range analysis is the \textit{(shallow) dimer} in a two-body system: when it exists, its binding energy is proportional to $\frac{1}{a^2}$.
 
Zero-range models are fully satisfactory for the two-body problem. In the center-of-mass frame, the system is described by the free Hamiltonian on $\rr^3$ away from the contact point, with the interaction encoded through the Bethe–Peierls boundary condition \cite{diplon}. If $\rho$ denotes the distance of the particle from the scattering center, then the wave function satisfies
\begin{equation} \label{BP_BC_two-body}
   \psi \sim  \frac{1}{\rho} - \frac{1}{a}, \quad \rho \rightarrow 0 \, .
\end{equation}
The situation is very different in the three-body sector. Imposing an analogous two-body contact condition leads to the so-called \enquote{Thomas effect} or \enquote{fall of the particles to the center} \cite{PhysRev.47.903,mf1,mf2}. In particular, the Ter-Martirosyan–Skornyakov (TMS) Hamiltonian for a three-boson system \cite{TMS} becomes unstable and its spectrum contains a sequence of eigenvalues unbounded from below. We also note that, in the three-body problem, the singular boundary condition is imposed on a three-dimensional submanifold rather than at isolated points. For this reason, to distinguish these conditions from the two-body Bethe–Peierls condition in \eqref{BP_BC_two-body}, we refer to them as \enquote{TMS boundary conditions} from now on (see Sec.~\ref{secII} for details).
 
It is well-known that the TMS Hamiltonian is symmetric but not self-adjoint, and therefore it does not define a valid quantum observable. Moreover, it was shown in \cite{Danilov,mf1,mf2} that every self-adjoint extension of this symmetric operator that  satisfies TMS boundary conditions analogous to \eqref{BP_BC_two-body} is unbounded from below. Minlos and Faddeev \cite{mf1} conjectured, however, that a suitable renormalization of the boundary condition leads to a lower-bounded zero-range Hamiltonian (see \eqref{TMS_bosonic_BC} and \eqref{TMS_fermionic_BC} below). This idea has been made rigorous recently in several works in mathematical physics, which construct and characterize the corresponding zero-range Hamiltonians (see, \emph{e.g.}, \cite{Basti2021ThreeBodyHW,Gallone2022SelfAdjointES,10.1007/978-981-99-5894-8_8,ferretti2024hamiltonianbosegas,ferretti2024hamiltonians}).
 
In the theoretical physics literature, to avoid the ultraviolet collapse present in the TMS Hamiltonian, authors introduce an additional short-distance input, such as considering a finite-range interaction or a short-range cutoff. This input may be represented by a \enquote{three-body parameter} that characterizes the Efimov spectrum. In this setting, the three-body parameter plays the role of a boundary condition at some short length scale $R_0$ for the three-body Schr\"odinger problem, fixed by a chosen observable. For example, one may identify it with the wavenumber of the first Efimov level $\kappa_*^{(1)}$ or the range of the interaction or derive an effective boundary condition at $R_0$ from a short-distance van der Waals potential (see, \emph{e.g.}, \cite{osti_4607608} and \cite{Review} for further discussion and references). As a result, \emph{universality} is often taken to refer mainly to the geometric law \eqref{universal_geometrical_law}, since the Efimov eigenvalues themselves depend on the chosen three-body parameter.
 
In summary, although zero-range theory provides an accurate description of the long-distance physics of three-body systems, the short-distance behavior encoded by contact conditions analogous to \eqref{BP_BC_two-body} must be regularized. In the case of finite-range potentials, regularization involves the explicit short-range details of the chosen potential, for example its range, effective range, cut-off, or van der Waals length. In a zero-range formulation, however, this information is not represented by a potential shape, but a three-body parameter still enters the problem through the choice of a scale or boundary condition.

Vitaly Efimov in his original work \cite{Efimov2} considered short-ranged potentials and, using hyperspherical coordinates, obtained the Efimov eigenvalues
\begin{equation}\label{Original_Efimov_eigenvalue_formula}
E_n=-\tfrac{1}{2 m_0 \, R_0^2} \, \mathrm{e}^{-2 \pi n /\left|s_0\right|} \exp \tfrac{2}{\left|s_0\right|}\left[\arctan \tfrac{\Lambda R_0}{\left|s_0\right|}-\Delta\right] \, .
\end{equation}
Here, $s_0$ is the parameter appearing in \eqref{universal_geometrical_law} and is determined by the hyperangular equation, $m_0$ is the appropriate reduced mass (which depends on the number of resonant pairs), $\Lambda$ plays the role of the three-body parameter (Efimov defined it as the logarithmic derivative of the interior solution at $R_0$, fixed by a momentum cutoff), and $\Delta$ is a phase that matches the solutions at $R_0$. More recently, Efimov’s original idea for three identical bosons has been given a rigorous formulation by Fermi, Ferretti, and Teta \cite{fermi_rigorous_2023}, who imposed a hard-core potential at the origin to avoid the Thomas effect.
 
It was later proposed (see, \emph{e.g.}, \cite{PhysRevLett.108.263001,PhysRevA.90.022106}) that, for three identical bosons, including a Lennard–Jones–type interaction $v_{\lambda_c} (r) := -\frac{C_6}{r^6} \left(1 - \frac{\lambda_c}{r^6}\right)$ in the pairwise potentials can make the three-body parameter universal, up to the van der Waals length $r_{\mathrm{vdW}} := \frac{1}{2} \left(\frac{2 \mu_{2b}C^6}{\hbar^2}\right)^{1/4}$. Here, $C_6$ is the species-dependent dispersion coefficient, $\lambda_c$ is a short-range cut-off, and $\mu_{2b}$ is two-body reduced mass \cite{RevModPhys.82.1225,PhysRevLett.108.263001}.
In hyperspherical coordinates, with hyperradius $R_{\mathrm{hyp}}$, the potential $v_{\lambda_c} (r)$ prevents the ultraviolet divergence by producing an effective repulsion barrier at $R_{\mathrm{hyp}} \sim 2 r_{\mathrm{vdW}}$. As a result, the three-body parameter, and in particular $\kappa_*^{(1)}$ is largely determined by the value of $r_{\mathrm{vdW}}$.
 
Wang et al. \cite{WangWang2012} extended this idea to the mass-imbalanced $2+1$ system consisting of two identical bosons and a third particle of different nature. In the Efimov-favored case of two heavy and one light particle, they analyze the problem within the Born--Oppenheimer approximation. In this approach, the heavy particles are treated as fixed scattering centers for the light particle: for each heavy–heavy separation one first solves the light-particle eigenvalue problem (see \eqref{fast} below), and then uses the corresponding eigenvalue as an effective potential in a reduced equation for heavy particles, leading to an effective Hamiltonian (see \eqref{slow} below).  The spectrum of this effective Hamiltonian provides an approximation to the three-body energies of the $2+1$ system (see Sec.~\ref{secII} for details).
 
Within this picture, imposing the boundary condition \eqref{BP_BC_two-body} on the light-particle Hamiltonian produces a short-distance singularity, analogous to the Thomas effect. More precisely, denoting by $r$ the separation between the two heavy particles, the resulting Born--Oppenheimer effective potential behaves as $-W^2(1)/r^2$ (see \emph{e.g.}, \cite{Albeverio} and Remark~\ref{local_remark} below). Here, $W$ is the Lambert-$W$ function, defined implicitly by $W(x) \ee^{W(x)}=x$; in particular, $W(1) \simeq 0.567143$ (see \cite{Corless:1996zz} for definitions and properties). As a consequence, the effective Hamiltonian is unstable and its spectrum is unbounded from below. Wang, \emph{et al.} showed that introducing a heavy–heavy interaction $v_{\lambda_c}(r)$ regularizes this short-distance behavior, allowing them to compute the Efimov spectrum and the corresponding wavenumbers $\kappa_*^{(n)}$.
 
In a closely related setting, Oi, Naidon, and Endo \cite{PhysRevA.110.033305} and Oi and Endo \cite{z5rx-51yx} studied the mass-imbalanced $2+1$ system of two identical fermions and a third particle within the Born--Oppenheimer approximation. In this case, fermionic statistics introduces a centrifugal repulsion in the heavy–heavy channel, which competes with the attractive effective potential. By including van der Waals and dipolar heavy–heavy interactions in the effective Hamiltonian, they showed the universality of the three-body parameter.
 
Prior to these studies, in 1979 Fonseca, Redish and Shanley analyzed the $2+1$ system for arbitrary angular momentum in \cite{Fonseca1} in the Born--Oppenheimer approximation. They assumed the light particle interacts with scattering centers via short-range Yamaguchi separable potential \cite{PhysRev.95.1628} of range $b$. At unitarity, they recovered an inverse-square tail in the Born--Oppenheimer effective potential and, as a consequence, established the existence of an Efimov spectrum. Away from the unitary limit, when the two-body scattering length is finite, they showed that the long-range behavior of the effective potential is of Yukawa form. In this regime the light particle localizes near one of the heavy scattering centers at the limit $ r \to \infty$, effectively reducing the three-body system to a dimer plus a spectator atom. They also computed the semi-classical Bargmann’s bound to show the logarithmic growth of number of bound states, and argued qualitatively that the spatial size of the weakest Efimov-like bound state is set by the two-body scattering length $a$. We note that the model studied by Fonseca et al. \cite{Fonseca1} does not include any additional heavy–heavy interaction.

These observations motivate us to revisit the zero-range description of the $2+1$ system within the Born--Oppenheimer approximation. In this setting, the light-particle Hamiltonian is described by a free particle in the presence of a two-center point interaction. Imposing the local contact condition \eqref{BP_BC_two-body} on this Hamiltonian leads to a pathological coincidence limit of the stationary scattering problem: as the distance between the two scattering centers tends to zero, the delta centers effectively disappear (non-additivity pathology), while the bound-state energy diverges to $-\infty$. This behavior is in contrast with the case of smooth short-range potentials (see, e.g., \cite{Mosta} and references therein from the theoretical physics literature, and \cite{FST,Albeverio,Fassari} in the mathematical physics literature). Consequently, the effective Hamiltonian is unstable, and the problem cannot be defined from the outset, similarly to the Thomas-effect singularity in the three-body system.

However, it is well-established in mathematical physics (though it may be less familiar in the theoretical physics literature) that, for two or more fixed centers, the Hamiltonians obtained by imposing (local) Bethe–Peierls conditions \eqref{BP_BC_two-body} form only a strict subclass of all possible point interactions (see \emph{e.g.}, \cite[Part II, Theorem 1.1.3 and subsequent remark]{Albeverio}). This is in contrast with the one-center case (which also describes the two-body problem in the center-of-mass frame) where the point interaction is fully characterized by the contact condition \eqref{BP_BC_two-body}. More importantly, we see that by characterizing the entire family of self-adjoint extensions, a large sub-class of point interactions corresponds to a well-defined stationary scattering problem, and its eigenvalue, which gives the effective potential for the heavy particles, remains regular at the origin. As a result, the Thomas-effect singularity is avoided. Physically, this can be viewed as a renormalization of the zero-range two-body interactions that depends on the configuration of all three particles. In this sense, it has an effect analogous to a three-body force, while preserving the point-like two-body nature of the interactions.
 
In our previous work \cite{FST}, following the 1985 construction of Dąbrowski and Grosse \cite{DG}, we focused on the non-additivity pathology of local point interactions. Dąbrowski and Grosse characterize the entire family of $n$-center point interactions in dimensions $d=1,2,3$ via von Neumann extension theory (see App.~\ref{App_construction} for a brief discussion and further references). We showed in \cite{FST} that a large sub-class of two-center Hamiltonians reduce to one-center Hamiltonians in the coincidence limit, provided the wave function is symmetric under exchange of the scattering centers, a setting appropriate for bosonic statistics.
 
As an application, we studied the $2+1$ mass-imbalanced system of two heavy bosons and a third particle in the $l=0$ channel. In this framework, the effective potential has the same value at the origin $r=0$ and $r \to \infty$, recovering the (shallow) dimer universality at the contact point of heavy particles. This, in turn, leads to universality of the Efimov spectrum at unitarity, up to the characteristic length of the problem. In other words, although the effective potential may be small but nonzero at intermediate and short distances, the long-range Efimov scaling is unchanged as long as the potential is sufficiently regular and tends to zero sufficiently fast as $r \to 0$.
 
In the present paper, we extend our analysis to antisymmetric exchange (fermionic case), and also higher angular-momentum channels. We also study the system away from unitarity. The analysis of the effective Hamiltonian \eqref{slow} for $l>0$, where the attractive Born--Oppenheimer potential competes with the repulsive centrifugal term, is technically more involved and requires tools beyond those used in the zero-angular-momentum sector.

To define a zero-range Hamiltonian as a physical model, one has to specify the length scale with respect to which the configuration variables are measured. This choice of scale is independent of whether the system is at unitarity or away from it. For example, if a heavy-heavy van der Waals interaction is included, the van der Waals length \(r_{\mathrm{vdW}}\) provides a natural length scale. For a finite-range short-range potential, the range of the potential can play this role. In a zero-range model, however, no finite interaction range is present.

In the two-body zero-range problem, the scattering length is the physical low-energy length scale. In particular, away from unitarity, the bound-state energy $-\frac{1}{a^2}$ is determined by the scattering length. At unitarity, the two-body zero-range problem becomes scale invariant. Thus the scattering length no longer provides a finite characteristic length, although the length scale used to measure the variables remains the same as the one chosen away from unitarity.



We set $L_0$ as the physical scale of the model and take it as the length unit throughout this paper. Let \(a_s\) denote the physical heavy-light scattering length measured experimentally. Then the reduced scattering length used in the rest of the paper is $a :=\frac{a_s}{L_0}$. Thus, all lengths appearing below are dimensionless unless stated otherwise. Physical lengths are recovered by multiplying the corresponding reduced quantities by $L_0$. We also note that $L_0$ may vary from system to system, just as microscopic length scales such as the van der Waals length $r_{vdW}$, or the range of a finite-range interaction, vary among different atomic mixtures.

Let $M$ and $m$ denote the heavy and light masses, respectively, and set the dimensionless parameter $\varepsilon^2 = \frac{4m}{2M + m}$. For angular momentum $l$, we show that, after fixing $L_0$, the Efimov spectrum has the form
\begin{equation}\label{universal_Efimov_true_scaling}
    \begin{split}
        &E_n = - \frac{4}{M\,r_0^2} \, e^{ \, \frac{2}{\beta} \left(   \arctan \left(\frac{2\beta}{2l+1} \right) + \phi_{\beta,0} -n \pi   \right)} \big(1 + \zeta_n \big) \\
        \text{with} \\
        &\phi_{\beta,0}= \arg \Gamma (1 + i \beta), \quad r_0 =\sqrt{2}\,\tfrac{3\pi}{4} L_0 \\
        &\beta = \sqrt{-l(l+1) + \ve^{-2} \,W^2(1) -1/4} \, .
    \end{split}
\end{equation}
Here, the error terms $\zeta_n$ tend to zero as $n \to \infty$. We observe that the ratio of successive bound states satisfies the geometrical law \eqref{universal_geometrical_law} with the parameter $\beta$.

In the formula \eqref{universal_Efimov_true_scaling}, the parameter $r_0$ plays the role of the three-body parameter, while the value $\sqrt{2}\,\tfrac{3\pi}{4}$ is uniquely determined by the construction of the light-particle zero-range Hamiltonian through von Neumann's formula. This value should therefore be understood as a consequence of the particular short-distance prescription adopted here, rather than as a universal microscopic constant. At the end of Sec.~\ref{secIII} and Remark~\ref{remark_theta_function}, we briefly discuss other possible constructions of the zero-range light-particle Hamiltonian, which may lead to different short-distance structures. This is consistent with the general role of regularization in contact three-body theories: the prescription is not meant to describe the microscopic interaction at arbitrarily small distances, but to remove the Thomas collapse and to fix the required three-body input. In what follows, in the rest of the paper, all lengths are measured in units of \(L_0\). Accordingly, we omit the explicit factor \(L_0\) from the notation, except for the numerical results in Sec.~\ref{secV}.

We stress that the present construction does not replace van der Waals universality as a microscopic explanation of the physical length scale in atomic systems. Rather, it provides an analytic zero-range counterpart: once the characteristic length $L_0$ is fixed, the three-body parameter is fixed by the construction and the Efimov spectrum follows explicitly. The price paid for this zero-range formulation is that, unlike in van der Waals universality, the model does not determine the absolute physical scale by itself; this scale must be supplied externally and adjusted to reproduce experimental data.

We remark that in \cite{Fonseca1}, and in much of the subsequent literature on the Born--Oppenheimer approach to Efimov physics, the three-body Hamiltonian is not introduced explicitly. Instead, one defines the light-particle Hamiltonian and the resulting effective Hamiltonian, and then assumes the validity of the Born--Oppenheimer picture to approximate the spectrum of the three-body system. In a zero-range setting, however, the very existence of a three-body Hamiltonian free of ultraviolet collapse is not trivial.
 
In this work, we do not construct the $2+1$ Hamiltonian in a fully rigorous way either; rather, we introduce it through renormalized TMS boundary conditions on the coincidence planes (see \eqref{TMS_bosonic_BC} and \eqref{TMS_fermionic_BC}). This viewpoint provides a direct link between the abstract construction of the light-particle Hamiltonian via von Neumann's formula and the underlying physical picture. It also gives a natural framework for discussing the validity and limitations of the Born--Oppenheimer approximation in Efimov physics (see Sec.~\ref{secVI}).
 
The structure of the paper is as outlined below: In Sec.~\ref{secII} we introduce the $2+1$ Hamiltonian and formalism of Born--Oppenheimer approximation. In Sec.~\ref{secIII} we construct the self-adjoint Hamiltonian of the light particle and explain why this zero-range model is physically relevant. In Sec.~\ref{secIV} we analyze the spectrum of the effective Hamiltonian, derive the Efimov eigenvalues \eqref{universal_Efimov_true_scaling} in the unitary limit, and discuss the spectrum away from unitarity: we estimate the spatial size of the weakest near-threshold state, compute a Bargmann's bound on the number of bound states, and describe how the spectrum deforms when the geometrical law \eqref{universal_geometrical_law} is no longer satisfied. In Sec.~\ref{secV} we present numerical results for selected alkali mixtures and for mixtures of alkali atoms with a different particle that are relevant for experiments. Finally, we conclude this paper in Sec.~\ref{secVI} with remarks on the validity of the Born--Oppenheimer approximation in Efimov physics and possible perspectives. Part of these results has been reported in \cite{saberbaghi2025family}.
\section{The three-body Hamiltonian and Formalism of Born--Oppenheimer approximation}\label{secII}
Throughout this paper, we set $\hbar^2=1$. Vectors are denoted in boldface, for example $\x$, and their Euclidean norm by the corresponding non-bold symbol, $x := |\x|$. For a (possibly unbounded) operator $A$, we write $D(A)$ for its domain. For $z \in \cc$, we denote its complex conjugate by $z^*$. We also use the notation $\langle \cdot , \cdot \rangle$ for the $L^2$ inner product. Moreover, we use $[L]$ and $[M]$ to denote the units of length and mass, respectively. Finally, we introduce the parameter $\alpha$, which is proportional to the inverse of the two-body scattering length, by
\begin{equation*}
   4 \pi \alpha := -\frac{1}{a} \, .
\end{equation*}
Throughout the model construction, lengths are expressed in units of a characteristic length $L_0$, supplied externally. We consider two identical heavy particles of mass $M$ at positions $\y_1,\y_2 \in \mathbb{R}^3$ and a light particle of mass $m$ at $\y_3 \in \mathbb{R}^3$. The three-body wave function belongs to either the bosonic or fermionic subspace $L^2_{b/f}(\mathbb{R}^9)\subset L^2(\mathbb{R}^9)$, consisting of functions that are symmetric or antisymmetric under the exchange of $\y_1$ and $\y_2$. Also, heavy particles interact with the light particle via zero-range forces.
 
Introducing Jacobi coordinates
\begin{equation*}
    \x = \y_3 - \tfrac{1}{2}(\y_1 + \y_2), \quad
\R = \y_1 - \y_2 \, ,
\end{equation*}
the center-of-mass, and the effective mass parameter
\begin{equation*}
    \x_{\mathrm{cm}} = \frac{M(\y_1 + \y_2) + m \y_3}{2M + m}, \quad
\gamma = \frac{2Mm}{2M + m} \, ,
\end{equation*}
we denote the Hamiltonian with bosonic (resp.\ fermionic) symmetry by $\mathcal{H}^{b}_{2+1}$ (resp.\ $\mathcal{H}^{f}_{2+1}$) and define the free Hamiltonian
\begin{equation}\label{2+1_freeHam}
    \hh_{2+1}^0 := -\frac{1}{2(2M + m)} \Delta_{\x_{\mathrm{cm}}}
- \frac{1}{M} \Delta_{\R}
- \frac{1}{2\gamma} \Delta_{\x} \, .
\end{equation}
Bosonic (resp. fermionic) symmetry translates into
\begin{equation*}
    \begin{split}
        &\Psi(x_{\mathrm{cm}}, x, R) = \Psi(x_{\mathrm{cm}}, x, -R) \\
        \Big(\text{resp. } &\Psi(x_{\mathrm{cm}}, x, R) = -\Psi(x_{\mathrm{cm}}, x, -R)\Big)
    \end{split} \, .
\end{equation*}
Henceforth, we neglect center-of-mass kinetic energy. Define the coincidence hyperplanes
\[
\Pi_\pm= \{(\R,\x) \in \rr^6|\; \x = \pm \R/2\} \, .
\] 
Away from these planes the particles do not coincide, so the interaction is absent and the dynamics is free. Accordingly, the Hamiltonian $\hh^{b/f}_{2+1}$ acts as the free Hamiltonian $\hh_{2+1}^0$ everywhere except on $\Pi_\pm$, on the configuration space \emph{i.e.} $\rr^6 \setminus \Pi_\pm$. The effect of the zero-range interaction is encoded entirely through boundary conditions imposed as one approaches $\Pi_\pm$. We also set $r := |\R|$ for the heavy–heavy separation, introduce the mass parameter $\mathscr{m}^{-1} := \frac{1}{4M} + \frac{1}{2\gamma}$, and let $\xi$ denote a function associated with boundary charges, belonging to the domain of $\hh^{b/f}_{2+1}$. 
 
Near the hyperplanes $\Pi_\pm$, that is, when
\begin{equation*}
    |\x -\y_i| =|\x \pm\R/2| \to 0, \, \R \neq 0 ,\, i=1,2 \, ,
\end{equation*}
the wave function satisfies singular TMS boundary conditions. Explicitly, up to an $o(1)$ term, in the bosonic case,
\begin{equation}\label{TMS_bosonic_BC}
    \psi^b(\x, \R) = \xi(\pm \R) \left(\tfrac{1}{4 \pi \mathscr{m}^{-1}\left|\x - y_i \right|} + \alpha + \tfrac{\theta_{\mathrm{3b}}(r)}{4 \pi \mathscr{m}^{-1}r}\right)  \,, 
\end{equation}
while in the fermionic case,
\begin{equation}\label{TMS_fermionic_BC}
    \psi^f(\x, \R) =(-1)^{i+1} \,\xi(\pm \R) \left(\tfrac{1}{4 \pi \mathscr{m}^{-1}\left|\x - y_i \right|} + \alpha - \tfrac{\theta_{\mathrm{3b}}(r)}{4 \pi \mathscr{m}^{-1}r}\right) 
\end{equation}
Here, $\theta_{\mathrm{3b}}(r)$ can be chosen from broad class of functions (see \cite{Basti2021ThreeBodyHW}) to prevent the Thomas effect. The TMS Hamiltonian associated with the boundary conditions \eqref{TMS_bosonic_BC} or \eqref{TMS_fermionic_BC} is a symmetric operator with infinite-dimensional defect spaces (see \eqref{defect_spaces_def} for the definition). On the other hand, the two-center point interaction Hamiltonian has a two-dimensional defect space. In this case, under the imposed symmetry and other natural technical assumptions, von Neumann's formula leads to a unique function $g_\alpha$. We will discuss in Sec.~\ref{secVI} and Remark~\ref{remark_theta_function} the relation between $\theta_{\mathrm{3b}}(r)$ in \eqref{TMS_bosonic_BC}–\eqref{TMS_fermionic_BC} and the function $g_\al(r)$ in \eqref{3D_bos_Bethe_BC}–\eqref{3D_fer_Bethe_BC}. Finally, we remark that \eqref{TMS_bosonic_BC} agrees with the boundary conditions introduced in \cite{Basti2021ThreeBodyHW}; note that the authors of \cite{Basti2021ThreeBodyHW} take identical bosons of mass $1/2$, so the corresponding mass parameter does not appear there.
 
Inspired by electromagnetism, the function $\xi$ is referred to as \enquote{charges} carried by the three-dimensional coincidence planes $\Pi_\pm$ in the literature. In physical variables its dimension is $[M]^{-1}[L]^{-1/2}$. In the reduced units used in this paper, the length dimension is absorbed into the scale $L_0$. For fixed $\R$, the TMS boundary conditions \eqref{TMS_bosonic_BC} and \eqref{TMS_fermionic_BC} reduce to the usual Bethe–Peierls form \eqref{BP_BC_two-body}, up to the renormalization term $\pm \frac{\theta(r)}{4 \pi \mathscr{m}^{-1}r}$.
 
We also note that the repulsive centrifugal term can stabilize the $2+1$ Hamiltonian even without the renormalization term $\pm \frac{\theta(r)}{4 \pi \mathscr{m}^{-1}r}$ in channels with angular momentum $l \geq 1$, provided the heavy–light mass ratio exceeds a critical value. For instance, in the $l=1$ channel, the Thomas effect is absent already without this renormalization when $\frac{M}{m}> 13.6$ (see \emph{e.g.}, \cite{BRAATEN2006259,doi:10.1142/S0129055X12500171,BeckerMichelangeliOttolini2018}). Nevertheless, including $\pm \frac{\theta(r)}{4 \pi \mathscr{m}^{-1}r}$ yields a Hamiltonian that is bounded from below for any mass ratio $\frac{M}{m}>1$. The mechanism behind this becomes clearer once we examine the structure of the $2+1$ problem and the role of angular momentum.
 
It is well-known that a zero-range heavy–light interaction couples only to the two-body $S$-wave channel (see \emph{e.g.}, \cite[Ch. X]{reed_methods_1975} and also, the treatment in \cite[Sec.~4.1]{Review}). This, however, should not be confused with the three-body (orbital) angular momentum of the $2+1$ system: after introducing Jacobi coordinate, the heavy–heavy Laplacian $-\Delta_{\R}$ still contains the usual angular part, which produces a repulsive centrifugal term in channels with $l \geq 1$.
 
If one imposes the unrenormalized TMS boundary conditions
\eqref{TMS_bosonic_BC} and \eqref{TMS_fermionic_BC}, the effective
short-distance behavior is governed by an attractive inverse-square term. In the \(l\)-th angular momentum channel, this term becomes supercritical whenever $\frac{M}{m}>\mathcal M_l^*$, and the corresponding Hamiltonian develops an ultraviolet collapse unless a
renormalization is introduced. Note that the same threshold \(\mathcal M_l^*\) is the critical mass ratio for the onset of the Efimov effect (see \eqref{critical_mass} and the
following remark).

The instability is not related to binding in the two-body heavy–light subsystems; it persists even at unitarity, where the pairs are at the zero-energy threshold, and drives the spectrum to $-\infty$ as the coincidence planes are approached. This signals the need for a theoretical short-distance adjustment rather than a physical prediction, and motivates imposing the renormalized TMS boundary conditions in all angular-momentum channels, even when centrifugal repulsion is present. Interestingly, in the Born--Oppenheimer formulation, this regularization is built into the light-particle Hamiltonian by default (see \eqref{3D_bos_Bethe_BC}–\eqref{3D_fer_Bethe_BC} below), while the centrifugal term enters only through the effective Hamiltonian.
 
Next we introduce the Born--Oppenheimer formalism. In this framework one exploits the separation of scales between the heavy and light particles: in the limit $\frac{M}{m} \to \infty$, the heavy particles move slowly compared with the light one and may be treated as approximately fixed while solving the light-particle problem. Simon \cite{Simon1983} studied the spectrum of systems similar to the effective Hamiltonian in the Born--Oppenheimer picture. Motivated by this viewpoint, we rescale $\mathcal{H}^{b/f}_{2+1}$ by the factor $2 \gamma$ (note that $ \gamma \to m$ as $M \to \infty$) and introduce the small parameter
\[
\varepsilon^2 = \frac{2\gamma}{M} = \frac{4m}{2M + m} \, ,
\]
After removing the center-of-mass motion, the free Hamiltonian $\hh^0_{2+1}$~\eqref{2+1_freeHam} in this scaling reads as:
\begin{equation}\label{free_3body_Ham}
    \hh_{\ve}^0 := 
- \ve^2 \Delta_{\R}
- \Delta_{\x} \, .
\end{equation}
Henceforth we work with the rescaled Hamiltonian $\hh^{b/f}_{\ve}$; results for $\hh^{b/f}_{2+1}$ follow by undoing the scaling. In particular, the Efimov spectrum \eqref{universal_Efimov_true_scaling} stated in the Introduction is expressed for $\hh^{b/f}_{2+1}$.
 
In the Born--Oppenheimer picture, the total wave function is factorized as
\begin{equation}
\Psi (\x, \R) = \psi (\x; \R)\, \Phi (\R) \nonumber
\end{equation}
Treating~$\R$ as a parameter, we (formally) define the light-particle Hamiltonian $\hh_{\al}^{b/f} $ by
\begin{equation} \label{fast}
 \left[ -\Delta_{\x} - \mu\delta(\x \pm \R/2) \right] \psi (\x; \R) = V_{\mathrm{eff}}(r)\, \psi (\x; \R)\, .
\end{equation}
Here, $\mu$ is coupling constant and Dirac deltas (with a great risk of confusion) formally represent point interactions with fixed scattering centers at $\pm \R/2$. In the reduced units of this paper, lengths are measured in units of $L_0$. Thus the powers of $L_0$ carried by the three-dimensional delta are absorbed into the rescaled coupling constant (Without the length rescaling, $\mu$ would have dimension $[L][M]^{-1}$). Also, it should not be interpreted as a bare physical coupling: after renormalization (see below) it is related to the extension parameter \(\alpha\), through the subtraction relation \(\mu^{-1}=\alpha+G_0(0)\). Since this formal notation is common in the physics literature but can be misleading if taken literally, we briefly recall how it should be interpreted.
 
For simplicity, consider the one-center formal operator $-\Delta_{\x} - \mu \delta(\x-\y)$, where $\y \in \rr^3$ is fixed. Let $G_{z}$ be the Green’s function of the Helmholtz operator, $\left(-\Delta - z\right) {G}_{z}(\x)  = \delta (\x)$, given explicitly in \eqref{Green's_function}. It is well known (see, e.g., \cite{Albeverio,reed_methods_1975,flamand1967applications}) that there is no local potential $V$ to substitute point interaction in $-\Delta_{\x} - \mu \delta(\x-\y)$. Nevertheless, a formal resolvent expansion with a lengthy but straightforward computation (see \cite[Ch.~II.1.1, Eqs.~(1.1.3)–(1.1.5)]{Albeverio}) leads to the renormalized identification $\mu^{-1} = \alpha + G_0(0)$. Here, $G_0(0)$ is divergent and must be understood through a subtraction procedure. Concretely, for $z=z^*= -\lambda, \, \lambda\in \rr^+$ and $G_\lambda := G_{z=-\lambda}$, one uses $\displaystyle\lim_{\x \to 0} G_0(\x) - G_\lambda(\x) = \frac{\sqrt{\lambda}}{4 \pi} $. As a consequence, for $\alpha <0$, the bound-state energy is $-\lambda = -(4 \pi \alpha)^2 = -\frac{1}{a^2}$, which is precisely the shallow-dimer universality in the two-body problem. For a rigorous construction via self-adjoint extension theory (von Neumann’s method), we refer to Appendix~\ref{App_construction}, where we also connect this characterization to the boundary condition \eqref{BP_BC_two-body}.
 
Returning to the Born--Oppenheimer formalism, we use the 
light-particle eigenvalue~$V_{\text{eff}}(r)$ as an effective potential for the heavy degrees of freedom. In other words, we replace the light-particle contribution in $\hh_\ve^{b/f}$ by $V_{\text{eff}}(r)$. This reduces the original three-body problem to an effective equation for the heavy-particle wave function:
\begin{equation} \label{slow}
\hh_{\mathrm{eff}}\, \Phi (\R) := \left[ - \ve^2 \Delta_{\R} + V_{\mathrm{eff}}(r) \right] \Phi (\R) = E^{\mathrm{eff}}\, \Phi (R) \, .
\end{equation}
Within the Born–Oppenheimer approximation, $E^{\mathrm{eff}}$ is taken as an approximation to the corresponding three-body energy of $\hh^{b/f}_{\ve}$.

The Born–Oppenheimer approach is classical in molecular physics (see, e.g., \cite{JeckoThierry}) and has also been used in Efimov physics for mass-imbalanced systems \cite{Fonseca1,Fonseca2,PhysRevA.80.022714,WangWang2012,Rosa_2019,PhysRevA.110.033305,FST,z5rx-51yx}. In general, its implementation for a $2+1$ system consists of three steps:
\begin{itemize}
    \item[1.] Construct and analyze the light-particle Hamiltonian \eqref{fast} in order to determine $V_{\mathrm{eff}} (r)$.
    \item[2.] Solve the effective heavy-particle problem \eqref{slow} to obtain $E^{\mathrm{eff}}$.
    \item[3.] Quantify the approximation error by comparing $E^{\mathrm{eff}}$ with the true three-body spectrum of $\hh^{b/f}_{\ve}$.
\end{itemize}
 In this paper we carry out the first two steps and comment on the validity of the approximation in the conclusion.
\section{Hamiltonian of the light particle}\label{secIII}
The Hamiltonian $\hh^b_\alpha$ with the formal expression \eqref{fast} and two point interactions was constructed in our previous work \cite{FST} using von Neumann’s extension theory, under the assumption of exchange symmetry between the scattering centers, which is the natural setting for bosonic statistics. Here, we briefly recall that construction and extend it to the antisymmetric (fermionic) case $\hh^{f}_\alpha$.

As discussed in the Introduction, a well-defined zero-range description of the $2+1$ quantum system or its Born--Oppenheimer approximation is beyond imposing the local boundary condition \eqref{BP_BC_two-body}. As a consequence, the construction of self-adjoint extensions of a densely defined symmetric operator and its technicality is unavoidable.

In this section we view $\hh^{b/f}_\alpha$ as the perturbation of the free Laplacian, supported on points $\y_1, \y_2$: 
\begin{equation*}
    \dot{\hh}_{0} = -\Delta_{\x} \restriction C^{\infty}_{0}(\mathbb{R}^3 \setminus \{\y_1,\y_2\}) \, .
\end{equation*}
The operator $ \dot{\hh}_{0}$ is symmetric and acts as the free Laplacian on $\rr^3$ away from the fixed scattering centers $\y_1, \y_2$. However, it is not self-adjoint, \emph{i.e.} the domain of $\hh^{b/f}_\alpha$ is strictly larger than $ D(\dot{\hh}_{0})$. In particular, besides regular $H^2$ functions, it contains singular components such as the Green’s function \eqref{Green's_function}.

The choice of domain for an unbounded operator can drastically change its spectral properties. A simple standard example is the one-dimensional momentum operator $-i \pd{}{x} f$ on $f \in C^1[0,2\pi]$, describing a particle on a finite interval. Imposing the quasi-periodic boundary condition $f(0) = \ee^{i \theta} f(2 \pi), \,0\leq\theta<2 \pi$, one obtains a symmetric operator whose closure is self-adjoint on $H^1[0,2\pi]$, with eigenvalues $n - \frac{\theta}{2\pi}, \, n \in \mathbb{Z}$. By contrast, imposing Dirichlet conditions $f(0) = f(2 \pi) = 0$, yields an operator with empty point spectrum. (see \emph{e.g.}, \cite{reed_methods_1975,Teschl2009MMQM}).

Also, unitary time evolution $\ee^{it A}$ exists precisely when the operator $A$ is densely defined and self-adjoint \cite{Teschl2009MMQM}. Thus, to obtain a valid quantum observable, one must construct self-adjoint extensions of the underlying symmetric operator.

In Appendix~\ref{App_construction} we briefly review von Neumann’s construction in the simpler one-center case, and then collect the detail of the characterization of $\hh^{b/f}_\alpha$ as the self-adjoint extension of $\dot{\hh}_{0}$. Specifically, we calculate the matrix $\Gamma^{b/f}_\alpha (-\lambda)$ in \eqref{resolvent2} below.

We use the resolvent formalism. The resolvent set is defined by $z \in \cc \setminus \sigma(\hh^{b/f}_\alpha)$, where $\sigma(\hh^{b/f}_\alpha)$ denotes the spectrum. For simplicity, we restrict to real spectral parameters $z = z^* = -\lambda, \lambda >0$, assuming $-\lambda$ is not an eigenvalue of $\hh^{b/f}_\alpha$. For $\x, \x' \in \rr^3$, the resolvent acting on $f \in L^2(\rr^3)$ has the form
\begin{widetext}
\begin{equation} \label{resolvent2}
 \left[ (\hh^{b/f}_\alpha + \lambda)^{-1} f\right] (\x) = \int_{\rr^3} G_\lambda (\x - \x')f(\x') \, \dd \x'
    +  \sum_{m,n =1}^{2}  \bigl[\Gamma^{b/f}_\alpha (-\lambda)^{-1}\bigr]_{mn}
       G_\lambda(\x-\y_m) \int_{\rr^3} G_\lambda(\y_n - \x') f(\x') \, \dd \x' \, .
\end{equation}
\end{widetext}
Here, the matrix $\Gamma^{b/f}_\alpha (-\lambda)$ is defined as
    \begin{equation} \label{Gamma-nonlocal}
[\Gamma^{b/f}_{\alpha}(-\lambda)]_{mn}= \left[ \frac{\sqrt{\lambda}}{4\pi} +\alpha \right] \delta_{mn} \pm \left[\frac{g_\alpha(r)}{4\pi r} - \frac{\ee^{-\sqrt{\lambda}r}}{4 \pi r}\right] (1 -
\delta_{mn})
\end{equation}
where \(+\) (resp. \( - \)) corresponds to exchange symmetry $\hh^{b}_\alpha$ (resp. exchange anti–symmetry $\hh^{f}_\alpha$) case. Also,
\begin{equation} \label{renor_func}
    g_\alpha(r) := \mathrm{e}^{-r/\sqrt{2}}\left( (4\sqrt{2}\pi\alpha +1)\sin\tfrac{r}{\sqrt{2}} + \cos\tfrac{r}{\sqrt{2}} \right).
\end{equation}
The function \(g_\alpha\) arises from the von Neumann construction (see \eqref{Matrix-Gamma} and \eqref{matrix_S_entries}).
 
For later use, we rewrite \eqref{resolvent2} more compactly as
\begin{equation}\label{resolvent_compact_form}
    (G_\lambda f)(\x) +  \sum_{m,n =1}^{2}  \bigl[\Gamma^{b/f}_\alpha (-\lambda)^{-1}\bigr]_{mn} G_\lambda(\x-\y_m)(G_\lambda f)(\y_n) \, .
\end{equation}
From this resolvent representation one recovers Bethe–Peierls–type boundary conditions near each center. We set $q\in\mathbb{C}$ the (point) charge, analogous to the boundary charge $\xi$ in \eqref{TMS_bosonic_BC}–\eqref{TMS_fermionic_BC}. For $|\x-\y_i|\to 0, \, i=1,2$ in the exchange symmetry case,
\begin{equation}\label{3D_bos_Bethe_BC}
\psi^{b}(\x)
=
q\left(\frac{1}{4\pi|\x-\y_i|}+\alpha+\frac{g_\alpha(r)}{4\pi r}\right)
+o(1) \, ,
\end{equation}
while in the antisymmetric case
\begin{equation}\label{3D_fer_Bethe_BC}
\psi^{f}(\x)
=
(-1)^{i+1}q\left(\frac{1}{4\pi|\x-\y_i|}+\alpha-\frac{g_\alpha(r)}{4\pi r}\right)
+o(1) \, .
\end{equation}
The boundary conditions \eqref{3D_bos_Bethe_BC}–\eqref{3D_fer_Bethe_BC} reduce to the local Bethe–Peierls form \eqref{BP_BC_two-body}, except for the additional term involving $g_\alpha(r)$. If one omits $\frac{g_\alpha(r)}{4 \pi r}$ and enforces only the local condition \eqref{BP_BC_two-body} at fixed separation $r$, one obtains a point-interaction model that suffers from several inconsistencies: in particular, the resolvent tends to the free resolvent, the scattering length of the two-center system (which differs from the two-body scattering length; see Corollary~\ref{corollary_scattering_length}) tends to zero as the centers merge. Also, the effective Hamiltonian in the Born–Oppenheimer approximation becomes unstable. These issues are avoided by characterizing the entire family of self-adjoint extensions and introducing the Hamiltonian $\mathcal{H}^{b/f}_{\alpha}$.  We summarize the main properties of $\mathcal{H}^{b/f}_{\alpha}$ in the following proposition.
\begin{proposition}\label{proposition3D}
\textbf{Case (i): Exchange symmetry.}  
\begin{enumerate}
    \item[1.] If  $r \to 0$, the resolvent \eqref{resolvent2} of $\mathcal{H}^b_\alpha$ converges in the strong resolvent sense to the resolvent of a one-center point interaction. Explicitly, for $f \in L^2(\rr^3)$ and setting $\y_1 = \y_2 = \y$, this resolvent reads as:
    \begin{equation} \label{two-becomes-one}
        (G_\lambda f)(\x) + \frac{4\pi}{\sqrt{\lambda} + 4 \pi \alpha}\, G_\lambda(\x-\y)\,
        G_\lambda f(\y).
    \end{equation}
    
    \item[2.] The operator $\mathcal{H}^b_\alpha$ has two eigenvalues, given as the solutions of the equation $\det \Gamma^b_\alpha(-\lambda) = 0$. They correspond to the eigenfunctions $G_{\lambda}(\x-\y_1) \pm G_{\lambda}(\x-\y_2)$ with the associated eigenvalues (in the implicit form)
\begin{equation*}
    \pm \mathrm{e}^{-\sqrt{\lambda}\,r} - \sqrt{\lambda}\,r -4 \pi \alpha r \mp g_\alpha(r) = 0 \, ,
\end{equation*}
where $g_\alpha(r)$ is defined in \eqref{renor_func}.

    \item[3.] For $\kk \in \mathbb{R}^3$ and $\x \in \rr^3 \setminus\{\y_1,\y_2\}\,$, generalized eigenfunctions of $\mathcal{H}^b_\alpha$ characterized as:
    \begin{equation} \label{generalized_eigenfunction_bosonic}
       \begin{split}
            &\Psi^b_{\alpha , \y_1, \y_2}(\kk , \x)
        = e^{i\kk \cdot \x} \\
        &+ \sum_{m,n=1}^{2}\left[\Gamma^b_\alpha (k^2)\right]_{mn}^{-1}
        e^{i\kk \cdot \y_{n}} \frac{e^{ik|\x-\y_{m}|}}{4 \pi |\x-\y_{m}|}.
       \end{split}
    \end{equation}
\end{enumerate}
\textbf{Case (ii): Exchange anti-symmetry.}
\begin{enumerate}
    \item[1.] If $r \to 0$, the resolvent \eqref{resolvent2} of $\mathcal{H}^f_\alpha$ converges in the strong resolvent sense to the free resolvent.
    
    \item[2.] The operator $\mathcal{H}^f_\alpha$ has two eigenvalues, given as the solutions of the equation $\det \Gamma^f_\alpha(-\lambda) = 0$. They correspond to the eigenfunctions $G_{\lambda}(\x-\y_1) \pm G_{\lambda}(\x-\y_2)$ with the associated eigenvalues (in the implicit form)
\begin{equation*}
     \mp \mathrm{e}^{-\sqrt{\lambda}\,r} - \sqrt{\lambda}\,r -4 \pi \alpha r \pm g_\alpha(r) = 0
\end{equation*}
where $g_\alpha(r)$ is defined in \eqref{renor_func}.

    \item[3.] For $\kk \in \mathbb{R}^3$ and $\x \in \rr^3 \setminus\{\y_1,\y_2\}\,$, generalized eigenfunctions of $\mathcal{H}^f_\alpha$ characterized as:
    \begin{equation} \label{generalized_eigenfunction_fermionic}
        \begin{split}
            &\Psi^f_{\alpha , \y_1, \y_2}(\kk , \x)
        = \ee^{i\kk \cdot \x} \\
        &+ \sum_{m,n=1}^{2} (-1)^{m+n}
        \left[\Gamma^f_\alpha (k^2)\right]_{mn}^{-1}
        \ee^{i\kk \cdot \y_{n}} \frac{\ee^{ik|\x-\y_{m}|}}{4 \pi |\x-\y_{m}|}.
        \end{split}
    \end{equation}
\end{enumerate}
\end{proposition}
Let us briefly comment on the above proposition.
\begin{itemize}
      \item[-] In the limit $r\to 0$ for the bosonic case, the limiting point interaction Hamiltonian has the binding energy $-(4\pi\alpha)^2 = -\frac{1}{a^2}$ for $\alpha <0$, \emph{i.e.} we recover the (shallow) dimer universality. Note that, the family \(\mathcal{H}^{b/f}_\alpha\) is a rank-two perturbation of the free Laplacian. Therefore, the strong resolvent convergence in the limit \(r\to 0\) cannot be improved to norm–resolvent convergence (see 
      \emph{e.g.} \cite[Theorem VIII.23]{reed1972functional}). 
      
      \item[-] Imposing anti-symmetry under exchange of the scattering centers, replaces \(\Gamma^b_\alpha(-\lambda)\) by \(\Gamma^f_\alpha(-\lambda)\), which differs only by a sign in the off-diagonal entries. In either symmetry case, only one eigenvalue equation is compatible with the symmetry, namely
      \begin{equation} \label{eig_eq}
          \mathrm{e}^{-\sqrt{\lambda}\,r} - \sqrt{\lambda}\,r - 4 \pi \alpha r- g_\alpha(r) = 0 \, .
      \end{equation}
      The corresponding eigenfunction is proportional to $G_{\lambda}(\x-\y_1) + G_{\lambda}(\x-\y_2)$ (resp. $G_{\lambda}(\x-\y_1) - G_{\lambda}(\x-\y_2)$) in the bosonic (resp. fermionic) case. Moreover, in \eqref{generalized_eigenfunction_fermionic} there is an extra factor \((-1)^{m+n}\) compared to \eqref{generalized_eigenfunction_bosonic}. 
    This sign is precisely what enforces anti-symmetry under exchange of the scattering centers, \emph{i.e.}, contributions associated with indices of opposite parity (\emph{i.e.}, \(m+n\) odd) pick up a minus sign. Additionally, both in the bosonic and fermionic cases, for $\kk\in\mathbb{R}^{3},\; \x\in\mathbb{R}^3\setminus\{\y_1,\y_2\}$, the generalized eigenfunctions (which are not square-integrable) satisfy 
    \[
    -\Delta\,\Psi^{b/f}_{\alpha , \y_1, \y_2}(\kk , \x)=k^2\,\Psi^{b/f}_{\alpha , \y_1, \y_2}(\kk , \x) .
    \]
\end{itemize}
By characterizing the generalized eigenfunctions of $\mathcal{H}^{b/f}_\alpha$, we can compute the (on-shell) scattering amplitude and, in particular, the scattering length.
Note that the standard two-body scattering length is defined for a single, radial potential and represents the \emph{effective length} that characterizes how the potential acts on a free quantum particle in low energies. However, in the present two-center setting, the interaction is neither radial nor reducible to a single
two-body channel; nevertheless, the scattering amplitude is well defined (see \cite[Section II.2.5]{Albeverio}). Denoting the unit sphere in three dimensions by $S^2$,
for directions $\omega, \omega^{\prime} \in S^2$ and $ k := |\kk| \geq 0$, the scattering amplitude reads:
\begin{equation}
\begin{aligned}
\mathscr{f}_{\alpha, \y_1, \y_2}&\left(k, \omega, \omega^{\prime}\right) \\  &=\lim _{\substack{|\x| \rightarrow \infty \\
\x /|\x|=\omega}}|\x| \mathrm{e}^{-i k|\x|}\left[\Psi^{b/f}_{\alpha, \y_1, \y_2}\left(k \omega^{\prime}, \x\right)-\mathrm{e}^{i k \omega^{\prime} \cdot \x}\right] \\
&=\frac{1}{4 \pi} \sum_{m, n=1}^2\left[\Gamma^b_\alpha\left(k^2\right)\right]_{m n}^{-1} \mathrm{e}^{i k\left(\y_m \cdot \omega^{\prime}-\y_n \cdot\omega\right)} \\
&=\frac{1}{4 \pi} \sum_{m, n=1}^2(-1)^{m+n}\left[\Gamma^f_\alpha\left(k^2\right)\right]_{m n}^{-1} \mathrm{e}^{i k\left(\y_m \cdot \omega^{\prime}-\y_n \cdot \omega\right)} \, .
\end{aligned}
\end{equation}
The scattering length of \(\mathcal{H}^{b/f}_\alpha\) is then defined as the zero-energy limit of the scattering amplitude $\mathscr{f}_{\alpha, \y_1, \y_2}\left(k, \omega, \omega^{\prime}\right)$. To distinguish it from the two-body scattering length $a$, we denote it by $a_\alpha (r)$ and set:
\begin{equation}\label{zero_energy_scattering_amplitude}
    a_\alpha (r) := -\displaystyle \lim_{ k \to 0}\mathscr{f}_{\alpha, \y_1, \y_2}\left(k, \omega, \omega^{\prime}\right) \,  .
\end{equation}
\begin{corollary}\label{corollary_scattering_length}
    For fixed $\alpha$, the scattering length $a_\alpha (r)$ of $\hh^{b/f}_\alpha$ defined in \eqref{zero_energy_scattering_amplitude} is computed explicitly as:
    \begin{equation*}
        \begin{split}
            a_\alpha (r) &=  \frac{-1}{4 \pi}\lim_{k \to 0} \sum^{2}_{m,n =1} \left[\Gamma^{b}_\alpha (k^2)^{-1}\right]_{mn} \\ 
            &=\frac{-1}{4 \pi} \lim_{k \to 0} \sum^{2}_{m,n =1} (-1)^{m+n}\left[\Gamma^{f}_\alpha (k^2)^{-1}\right]_{mn} \\ &= \frac{2r}{-4\pi\alpha r - g_\alpha(r) + 1} \, ,
        \end{split}
    \end{equation*}
    where $g_\alpha(r)$ defined in \eqref{renor_func}. In particular:
    \begin{equation}\label{asym-scattering}
         a_\alpha (r) = \left\{ \begin{aligned}
            & \tfrac{ -2}{4 \pi \alpha}   \;\;\;    &\text{for} \quad & r\to +\infty\\
            &\tfrac{ -1}{4 \pi \alpha}     &\text{for} \quad & r \to 0
            \end{aligned}\;\right. .
    \end{equation}
\end{corollary}
Before discussing the physical relevance of the Hamiltonian $\hh^{b/f}_\alpha$, we first analyze the eigenvalue equation \eqref{eig_eq}, recalling that \(-\lambda(r)\) plays the role of the effective potential in the Born--Oppenheimer equation \eqref{slow}. The implicit equation \eqref{eig_eq} can be solved explicitly in terms of the Lambert \(W\)-function. It plays the role of the effective potential in \eqref{slow}, \emph{i.e.}, $V_{\mathrm{eff}}(r)$. For $\alpha \leq 0$, we obtain
\begin{equation}\label{E0}
     - \lambda_\alpha (r) = -\frac{1}{r^2} \left[W\left(\ee^{ 4 \pi \alpha r + g_\alpha(r)}\right) - \left(4 \pi \alpha r + g_\alpha(r)\right)\right]^2 
\end{equation}
where, $W$ denotes the principal branch of the Lambert function. For $\alpha >0 $, the spectrum of $\hh^{b/f}_\alpha$ has no discrete spectrum, so its spectrum consists only of the purely essential spectrum.
 
The function \(-\lambda_\alpha (r)\) in \eqref{E0} is analytic and in the unitary limit $\alpha = 0$ (infinite two-body scattering length), has the expected inverse-square behavior at large distances. To study the universality of the Efimov spectrum and its deformation away from unitarity, we need precise estimates on \(-\lambda_\alpha (r)\). The next lemma collects these properties; its proof is given in Appendix~\ref{appendix1}.
\begin{lemma}\label{analyticity_lemma}
    The function $-\lambda_\alpha (r)$ defined in \eqref{E0} is analytic for $r > 0$. Furthermore, for $\alpha = 0$, we have
\begin{equation}\label{E01}
-\lambda_0 (r) = \left\{ \begin{aligned}
          & -\frac{W^2(1)}{r^2} + O \big(\ee^{-r/\sqrt{2}} \big)              \;\;\;    &\text{for} \quad & r\to +\infty\\
         &- \frac{r^2}{16} + O\big( r^3\big)     &\text{for} \quad & r \to 0\,
        \end{aligned}\;\right. .
\end{equation}
Moreover, the singularity at \(r=0\) is removable, and the resulting analytic branch can be continued along the real half-line. Also, for $\alpha < 0$,
\begin{equation}\label{E02}
-\lambda_\alpha (r) = \left\{ \begin{aligned}
          &  -(4 \pi \alpha)^2 + O\Big( \frac{\ee^{-  r/\sqrt{2} }}{r} \Big)              \;\;\;    &\text{for} \quad & r\to +\infty\\
         & -(4 \pi \alpha)^2 + O(r)   &\text{for} \quad & r \to 0\,
        \end{aligned}\;\right. .
\end{equation}
The corresponding analytic branch can be continued along the interval \([a,\infty)\), where \(a=-1/(4\pi\alpha)\).
\end{lemma}
\begin{remark} \label{local_remark}
    For \emph{local} point interactions defined by imposing the contact condition \eqref{BP_BC_two-body} at each scattering center, the term $\frac{g_\alpha(r)}{4 \pi r}$ is absent from \eqref{Gamma-nonlocal} (see \cite[Chapter II.1, p.~119]{Albeverio}). Consequently, for $\alpha <0$, the analogue of \eqref{E0} is
    \begin{equation}\label{eig_local}
        -\lambda^{\mathrm{local}}_\alpha (r) = -\frac{1}{r^2} \left[W\left(\ee^{ 4 \pi \alpha r}\right) - 4 \pi \alpha r \right]^2 \, .
    \end{equation}
    Figure~\ref{Figure_localvsnon-local} compares $-\lambda^{\mathrm{local}}_\alpha$ and $-\lambda_\alpha$ in the unitary limit $\alpha =0$.
\end{remark}
Here, we constructed the Hamiltonian $\hh_\alpha^{b/f}$ via von Neumann's formula. We remark that this Hamiltonian can be rigorously approximated by separable potentials, provided that potentials are regular enough (For details, see \cite[Sec. IV]{DG}). At the same time, other renormalization schemes for (local) point interactions have been proposed in the literature (see, \emph{e.g.}, \cite{Mosta,Fassari,ferretti2024hamiltonians}, and also \cite{basti2026zero} for a recent contribution). 
In particular, Ferretti and Teta replace \(g_\alpha\) in \eqref{3D_bos_Bethe_BC} by a function \(\theta(r)\) such that \(\theta'(0)=0\) \cite[Proposition~4.2]{ferretti2024hamiltonians} (or more generally \(\theta'(0)=c\) for a constant $c$). It is easy to see that if $g_\alpha$ in the eigenvalue equation \eqref{E0} is replaced by a suitable function $\theta(r)$, then
        \begin{equation}\label{behav_theta_renorm}
            V^{\theta}_{\mathrm{eff}}(0)=-\frac{\left[4 \pi \alpha  + \theta^{\prime}(0)\right]^2}{4} \, .
        \end{equation}
In this case, although Proposition~\ref{proposition3D} remains valid in this case, the factor $\frac{4\pi}{\sqrt{\lambda} + 4 \pi \alpha}$ in \eqref{two-becomes-one} is replaced by 
\begin{equation*}
    \frac{4\pi}{\sqrt{\lambda} + \left(\frac{4 \pi \alpha  + \theta^{\prime}(0)}{2}\right)} \, .
\end{equation*}
Although the choice of the renormalization function is arbitrary, this raises the natural question of which model is physically relevant. We see, in fact, that the behavior of $g_\alpha$ at the origin plays a crucial role in having a satisfactory scattering problem. 

More precisely, when the two heavy particles are far apart, one expects the light particle to bind to either center, so that (away from unitarity) the system behaves as a dimer of energy $-\frac{1}{a^2}$ plus a free particle. This long-distance behavior does not depend on the chosen renormalization. In particular, it suggests that the scattering length associated with \eqref{fast} at $r \to \infty$ should be twice its value \(r\to 0\): at large separation the free particle has two binding options, whereas at the origin $r=0$, the two centers effectively act as a single one. This is exactly the behavior in \eqref{asym-scattering}, which amounts to the asymptotics of the effective potential in \eqref{E01} and \eqref{E02}.
\section{The spectrum of the effective Hamiltonian}\label{secIV}
In this section we study the eigenvalues of the Hamiltonian $\hh_{\mathrm{eff}}$ in \eqref{slow}, where the effective potential $V_{\mathrm{eff}}(r)$ is given by \eqref{E0}. By the Kato–Rellich theorem the operator is self-adjoint (see, \emph{e.g.}, \cite{ReSi}). Moreover, using \eqref{E01} and \eqref{E02}, we see that
\[
\sigma_{\mathrm{ess}}(\hh_{\mathrm{eff}})=\Bigl[-(4 \pi \alpha)^2,\,+\infty\Bigr),
\]
and any discrete eigenvalues satisfy
\[
\sigma_{\mathrm{p}}(\hh_{\mathrm{eff}})\subset \bigl[\min V_{\mathrm{eff}},\, -(4 \pi \alpha)^2\bigr).
\]
\\
Furthermore, since the potential \(V_{\mathrm{eff}}(r)\) is radial, \(D(\hh_{\mathrm{eff}})\) is invariant under rotations, and we may apply the partial-wave expansion. Fixing the angular momentum \(l \geq 0\), we decompose \(\Phi\in L^2(\mathbb{R}^3)\) into a radial part \(r^{-1}u(r)\in L^2(\mathbb{R}^+; r^2\,\mathrm{d}r)\) and spherical harmonics \(Y_{lm}(\phi,\theta)\in L^2(S^2;\mathrm{d}\Omega)\), where \(S^2\) is the unit sphere in \(\mathbb{R}^3\) with surface measure \(\mathrm{d}\Omega\). Applying the unitary map \(L^2(\mathbb{R}^+; r^2\,\mathrm{d}r)\to L^2(\mathbb{R}^+;\mathrm{d}r)\) given by \(f\mapsto r f\) yields the radial equation for \(u(r) \in L^2(\mathbb{R}^{+})\) :
\begin{equation}\label{reduced-radial-higher-momentum}
    \begin{split}
        -\ve^2&\left(u^{\prime \prime}(r) - \frac{l(l+1)}{r^2} u(r) \right) +V_{\mathrm{eff}}(r) u(r) = E^{\mathrm{eff}}u(r),    \\
        &\text{for} \,\,\, l=0, \quad  u(0)= 0 \\
        &\text{for} \,\,\, l \geq 1, \quad \lim_{r \to 0} u(r)= 0, \quad \lim_{r \to 0}r^{-(l+1)}u(r)=1
    \end{split}
\end{equation}
where \(u(r)=r\Phi\) (see \cite[Appendix to X.1, Example 4]{reed_methods_1975}, \cite[Theorem XI.53]{RS3}, \cite[Ch.\ XIII.3.B]{ReSi}). In general, \eqref{reduced-radial-higher-momentum} is understood as a differential operator acting on \(L^2(\mathbb{R}^+)\). However, since the effective potential is analytic (see Lemma~\ref{analyticity_lemma}), any solution of \eqref{reduced-radial-higher-momentum} is smooth. Henceforth, we focus our analysis on the Hamiltonian \eqref{reduced-radial-higher-momentum}.
\subsection{Efimov effect in the unitary limit}\label{sec:Efimov_unitary}
The main result of this subsection is the following theorem:
\begin{theorem}\label{main_theorem}
There exists an infinite sequence of negative eigenvalues $E^{\mathrm{eff}}_n$ of the problem \eqref{reduced-radial-higher-momentum} and the effective potential \eqref{E0} with $\alpha=0$, such that $E^{\mathrm{eff}}_n \rightarrow 0$ for $n \rightarrow \infty$. Moreover, 
\begin{equation}\label{universal_Efimov_eig}
 \begin{split}
        &E^{\mathrm{eff}}_n = -\ve^2 \frac{4}{r_0^2} \, e^{ \, \frac{2}{\beta} \left(   \arctan \left(\frac{2\beta}{2l+1} \right) + \phi_{\beta,0} -n \pi   \right)} \big(1 + \zeta_n \big) \\
        \text{with} \\
        &r_0 = \sqrt{2}\,\tfrac{3\pi}{4} , \quad \phi_{\beta,0}= \arg \Gamma (1 + i \beta), \\
        &\beta = \sqrt{-l(l+1) + \ve^{-2} \,W^2(1) -1/4} \, .
    \end{split}
\end{equation}
where 
$\;\zeta_n\rightarrow 0\,, \;\; \text{for}\;\; n \rightarrow \infty$.
\end{theorem}
\begin{corollary}\label{corollary_Efimov_geometrical_law}
  The eigenvalue formula \eqref{universal_Efimov_eig} yields the Efimov geometric law
\begin{equation}\label{BO_geometrical_law}
\lim_{n \rightarrow \infty} \frac{E^{\mathrm{eff}}_n}{E^{\mathrm{eff}}_{n+1}} = \ee^{\frac{2 \pi}{\beta}}\,. 
\end{equation}
\end{corollary}
The error terms $\zeta_n$ are $o(1)$ as $n \to \infty$. They arise from two sources: first, from the nonzero short-range part of the effective potential \eqref{E0} (see Lemma~\ref{lemma_Ulam} below); and second, from systematic corrections due to the fact that the modified Bessel function $K$ with imaginary index is the solution of \eqref{reduced-radial-higher-momentum} with the effective potential of form $-\frac{W^2(1)}{r^2}$ (see the proof of Proposition~\ref{proposition_Efimov_constant_potential}). Consequently, the geometric law \eqref{BO_geometrical_law} holds strictly only in the limit $n \to \infty$, and deviations are expected for the lowest Efimov levels. We compute the first-order correction in Lemma~\ref{Lemma_second_order_error}.
 
We prove Theorem~\ref{main_theorem} by solving the problem \eqref{reduced-radial-higher-momentum} with the simplified potential $V_{\mathrm{eff}}(r)=0$ for $r<r_0$ and $V_{\mathrm{eff}}(r)=-\frac{W^2(1)}{r^2}$ for $r \geq r_0$. Then, we extend the result to the full potential \(V_{\mathrm{eff}}(r)=-\lambda_0(r)\). For \(\alpha=0\), the latter is well approximated by $-\frac{W^2(1)}{r^2}$ for \(r>r_0\), up to exponentially decaying oscillations, with \(r_0 = \sqrt{2}\,\tfrac{3\pi}{4}\) (see  Figure~\ref{Figure_localvsnon-local} below).

We show that, although \(-\lambda_0(r)\) is nonzero yet small at short distances \(r<r_0\), the low-energy character of the Efimov spectrum makes this short-range behavior irrelevant, provided \(-\lambda_0(0)=0\).
\begin{figure}[b]
        \centering
        \includegraphics[width=9cm]{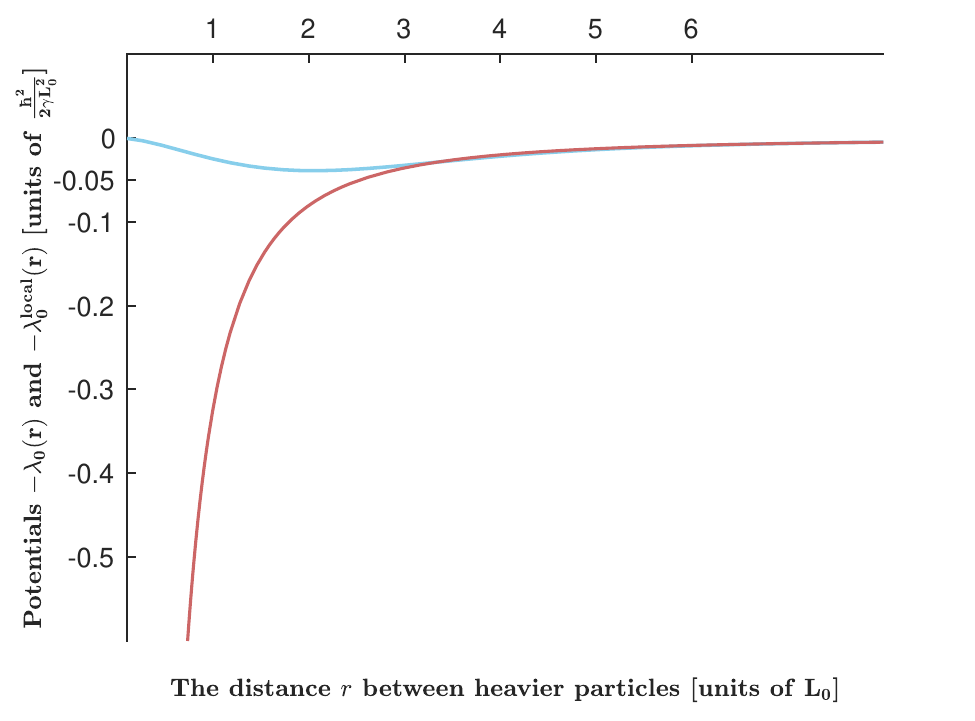}
        \caption{\footnotesize{The light blue curve represents $-\lambda_{0}(r)$ in \eqref{E0}, while the red curve is the graph of $\frac{-W^{2}(1)}{r^{2}}$, that is, $-\lambda^{\mathrm{local}}_{0}(r)$ in \eqref{eig_local}. The two curves intersect at $r_{k} = \sqrt{2}(\frac{3\pi}{4} + k \pi),\,k \in \mathbb{N}_0$, and, up to an exponentially decaying oscillation, they coincide for $r > r_{0} $.}}
        \label{Figure_localvsnon-local}
    \end{figure}
    
From \eqref{E0} , we have $- \lambda_0 (r) = -\frac{1}{r^2} \left[W\left(\ee^{g_\alpha(r)}\right) - g_\alpha(r)\right]^2$, where $g_\alpha(r)$ is defined in \eqref{renor_func}.  \(r_k = \sqrt{2}\left(\frac{3\pi}{4} + k \pi\right),\,\, k \in \mathbb{N}_0\), are roots of \(g_\alpha(r)\) for \(\alpha = 0\). We replace \(-\lambda_0(r)\) by the piecewise potential
\begin{equation}\label{k_potentials}
    V_{k}(r) =
    \begin{cases}
          -\lambda_0(r)   & \text{if } 0\leq r \leq r_k,\\[2mm]
         -\dfrac{W^2(1)}{r^2}  & \text{if } r>r_k \,.
    \end{cases}
\end{equation}
Moreover, for \(k \in \mathbb{N}_0\), define the operators \(\mathcal{H}_k := -\varepsilon^2 \Delta_{\R} + V_k\). It is straightforward to see that \(\mathcal{H}_k\) converges to \(\mathcal{H}_{\mathrm{eff}}\) in the norm-resolvent sense as \(k \to \infty\) (see \cite[proof of Theorem~4.4]{FST}). Therefore, once the Efimov effect is established for each \(\mathcal{H}_k\), it extends to the full Hamiltonian \(\mathcal{H}_{\mathrm{eff}}\).
 
Furthermore, it is well-known that the radial operator $-u^{\prime\prime}(r) - \frac{\mu}{r^2}u(r)$ has infinitely many eigenvalues accumulating at zero if and only if \(\mu>\tfrac{1}{4}\) (see, e.g., \cite[Sec.~4.6]{Sc}). In addition, for any angular momentum \(l\ge 1\), the centrifugal term \(\varepsilon^2 \tfrac{l(l+1)}{r^2}\) competes with the attractive part of \(V_k(r)\). More precisely, for each integer \(l\ge 1\), the problem \eqref{reduced-radial-higher-momentum} with an effective potential of the form \eqref{k_potentials} has infinitely many eigenvalues if and only if
\begin{equation}\label{critical_mass}
    \frac{M}{m} \,\ge\, \mathcal{M}_l^{*},
\qquad\text{where}\qquad
\mathcal{M}_l^{*} =\frac{4l(l+1) + 1 - W^2(1)}{2W^2(1)} \, .
\end{equation}
The value \(\mathcal{M}_l^{*}\) is close to the critical mass ratio obtained from a three-body analysis in hyperspherical coordinates. For instance, in the \(l=1\) channel we find \(\mathcal{M}_1^{*}\simeq 13.4902\), to be compared with the critical mass \(13.6069\) reported by the hyperspherical method (see, \emph{e.g.}, \cite[Sec.~6.2.4]{Review} and references therein). We also note that the critical mass ratio \eqref{critical_mass} differs slightly from the condition given in \cite[Eq.~(6.14)]{Review}, where the factor \(\tfrac{2m}{2m+1}\) was approximated by \(1\). We remark that the logarithmic growth in the number of eigenvalues in the channel $l=1$ was rigorously studied in \cite{Basti2017}.
 
Now, we prove the Efimov effect for \eqref{reduced-radial-higher-momentum} with an effective potential of the form
\begin{equation}  \label{auxiliary_potential}
    V_{\mathrm{aux}}(r) = \left\{ \begin{aligned}
          &\Lambda               && \text{if } 0\le r \le r_0,\\
         -&\dfrac{W^2(1)}{r^2}   && \text{if } r>r_0,
        \end{aligned}\right.
\end{equation}
where \(\Lambda \leq 0\) is a constant and \(r_0>0\) is arbitrary. We also rescale the energy in \eqref{reduced-radial-higher-momentum} by setting
\begin{equation} \label{scaling_energy}
    \begin{split}
        &E^{\mathrm{eff}} = -\ve^2 \eta^2, \qquad \eta>0,\\
        &\xi_0 = \sqrt{\ve^{-2}\Lambda}\, r_0 \, .
    \end{split}
\end{equation}
A special case of Proposition~\ref{proposition_Efimov_constant_potential} (with \(\Lambda=0\) and \(l=0\)) was proved in \cite[Prop.~4.1]{FST}. The present proof follows a similar strategy and is given in Appendix~\ref{appendix2}. We then extend Proposition~\ref{proposition_Efimov_constant_potential} to the potentials \eqref{k_potentials}, in particular to \(V_0\) of the form \eqref{k_potentials} with \(r_0=\sqrt{2}\,\frac{3\pi}{4}\).
\begin{proposition} \label{proposition_Efimov_constant_potential}
For the problem \eqref{reduced-radial-higher-momentum} with the potential \eqref{auxiliary_potential}, there exists an infinite sequence of negative eigenvalues \(E^{\mathrm{eff}}_n\) such that \(E^{\mathrm{eff}}_n \to 0\) as \(n\to\infty\). Moreover,
\begin{equation}\label{asei}
    E^{\mathrm{eff}}_n = -\ve^2\frac{4}{r_0^2} \, e^{ \, \frac{2}{\beta} \Big(   \arctan \big(2\beta f\left(\xi_0\right)\big) + \phi_{\beta,0} -n \pi   \Big)} \big(1 + \zeta_n \big)
\end{equation}
with $\beta$, $\phi_{\beta,0}$ as in \eqref{universal_Efimov_eig}. Also,  
$\;\zeta_n\rightarrow 0\,, \;\; \text{for}\;\; n \rightarrow \infty$, and
\begin{equation*}
    f(\xi_0) :=\frac{I_{l + 1/2} \left(\xi_0\right)}{2\xi_0 \, I_{l + 1/2}^{\prime} \left(\xi_0\right)} \, ,
\end{equation*}
where $I$ is the modified Bessel function of the first kind. In particular, for $\Lambda =0$ in \eqref{auxiliary_potential}, 
\begin{equation*}
    \lim_{\xi_0 \rightarrow 0} f(\xi_0)=\frac{1}{2l+1} \, .
\end{equation*}
\end{proposition}
\begin{remark}
We can also characterize the eigenfunctions for \eqref{reduced-radial-higher-momentum} with the potential \eqref{auxiliary_potential}. Recall that \eqref{reduced-radial-higher-momentum} was derived from \eqref{slow} by setting \(u(r)=r\,\Phi\). Using the solutions \eqref{solutions} in the proof of the proposition, the eigenfunction corresponding to \(E^{\mathrm{eff}}_n\) can be written as
\begin{widetext}
    \begin{equation}\label{wave_func_Efimov}
\Phi_n(r)= \left\{ 
\begin{aligned}
&D\, \tfrac{ \sqrt{r_0} \,K_{i\beta}(\eta_n r_0) }{ (\sqrt{\ve^{-2}\left|E^{\mathrm{eff}}_n\right|+\Lambda}\, r_0)^{\frac{1}{2}} I_{l + 1/2} (\sqrt{\ve^{-2}\left|E^{\mathrm{eff}}_n\right|+\Lambda}\, r_0)} \,
 \tfrac{(\sqrt{\ve^{-2}\left|E^{\mathrm{eff}}_n\right|+\Lambda}\, r)^{\frac{1}{2}} I_{l + 1/2} (\sqrt{\ve^{-2}\left|E^{\mathrm{eff}}_n\right|+\Lambda}\, r)}{r} \,,  &r<r_0\,\\
&D\, \tfrac{ K_{i\beta} (\sqrt{\ve^{-2}\left|E^{\mathrm{eff}}_n\right|}r)}{\sqrt{r}} \,,  &r>r_0\,
\end{aligned}
\right. 
\end{equation}
\end{widetext}
where $D$ is a normalization constant. 
\end{remark}
To show that, for $\Lambda=0$, the eigenvalue formula \eqref{asei} (with a possibly different sequence $\zeta_n$ still of order $o(1)$) also holds for the full effective potential $V_{\mathrm{eff}}(r)=-\lambda_0(r)$, we compare solutions of \eqref{reduced-radial-higher-momentum} for the auxiliary potential \eqref{auxiliary_potential} with those for $V_{\mathrm{eff}} = V_k$ in \eqref{k_potentials}. The goal is to control the latter by the former in the limit $\eta \to 0$. The value of the effective potential at the origin, $V_{\mathrm{eff}}(0)$, plays a crucial role in this argument.
 
In \cite{FST} we used an ad-hoc argument to prove Lemma~\ref{lemma_Ulam} for the special case $l=0$, which relied heavily on the positivity of $-\lambda^{\prime}_0(r)$ on the $(r_{\text{min}}, r_0]$, where $r_{\text{min}}$ is the point that $-\lambda_0(r)$ attains its minimum. For higher angular momenta this condition is no longer guaranteed, because the repulsive centrifugal term competes with the attractive potential.
\\
Our main tool is Ulam stability for second-order ordinary differential equations (see, \emph{e.g.}, \cite{ANDERSON2025128908, sym12091451}). A brief review of this topic and the proof of following lemma are given in Appendix~\ref{Appendix_Ulam}. 
\begin{lemma}\label{lemma_Ulam}
    Let $f(r)$ (resp. $f_0(r)$) denote the solution of \eqref{reduced-radial-higher-momentum} with effective potential $V_{\mathrm{eff}} = V_k$ in \eqref{k_potentials} (resp. $V_{\mathrm{eff}} = V_{\mathrm{aux}}$ in \eqref{auxiliary_potential}). Then, for any $l \geq 0$, on the interval $[0,r_k]$ with $r_k = \sqrt{2}(\frac{3\pi}{4} + k \pi),\,\, k \in \mathbb{N}_0$, there exists a constant $C_k>0$ such that
        \begin{equation}\label{Ulam_error}
            \big\lvert f(r) - f_0(r) \big\rvert \leq C_k \left(\eta r\right)^{l+1}
        \end{equation}
\end{lemma}

Finally, we can prove Theorem~\ref{main_theorem}. 

\begin{proof}[\textbf{Proof of Theorem~\ref{main_theorem}}] \label{proof_main_theorem}
    Let $E^{\mathrm{eff}}_n$ (resp. $E^0_n$) denote the $n$-th eigenvalue of the problem \eqref{reduced-radial-higher-momentum} with the effective potential $-\lambda_0(r)$ in \eqref{E0} (resp. $V_0$ in \eqref{k_potentials}). By min-max principle (see \emph{e.g.}, \cite{ReSi}), and the convergence of $\hh_k \to \hh_{\mathrm{eff}}$ in the norm-resolvent sense, we have $\displaystyle \lim_{n \to \infty} \big| E^{\mathrm{eff}}_n - E^0_n \big| \to 0$. Here, with abuse of notation, we absorbed this error into the systematic error $\zeta_n$ in \eqref{asei}. Putting together Proposition~\ref{proposition_Efimov_constant_potential}, Lemma~\ref{lemma_Ulam}, and the discussion above, we obtain the eigenvalue formula \eqref{universal_Efimov_eig}.
\end{proof}
\begin{remark}\label{remark_Ulam}
    The asymptotic behavior of $\lambda_0(r)$ at the origin is crucial for obtaining an Efimov spectrum of the form \eqref{universal_Efimov_eig}. More precisely, if the constant $\Lambda$ in \eqref{auxiliary_potential} is replaced by a regular function $\Lambda(r)$ with nonzero value at $r=0$, then an argument similar to Lemma~\ref{lemma_Ulam} no longer yields \eqref{asei}, because the error
    $(\sqrt{\ve^{-2}\Lambda}\, r_0)^{\frac{1}{2}}$ does not vanish in the limit $\eta \to 0$ (see \eqref{so<}). Therefore, Efimov spectrum depends on the choice of the renormalization function and its shape near the origin, even if different choices share the same value  $\Lambda$ at $r=0$. Intuitively, imposing a non-zero $\Lambda$ at the origin means that the three-body system generates an energy larger than the expected two-body value $-\frac{1}{a^2}$; this shift then enters the Efimov spectrum beyond the characteristic length, unless the short-range part is a sharp constant potential as in \eqref{auxiliary_potential}.
\end{remark}
We conclude this subsection with a brief discussion of possible non-Efimov eigenvalues of the Hamiltonian \eqref{reduced-radial-higher-momentum}. Loosely speaking, a non-Efimov eigenvalue (see, \emph{e.g.}, \cite{10.1063/1.2400657,PhysRevA.109.043323}) is an eigenvalue generated by the short-range part of the interaction and its spatial size is significantly smaller than that of Efimov states. Such eigenvalues do not obey the scaling law \eqref{universal_geometrical_law}. Consequently, if no non-Efimov eigenvalue is present, the ground state of the system is the first Efimov eigenvalue.

Let $r_0 = \sqrt{2} \frac{3 \pi}{4} $ and set $\Lambda = -\lambda_0(r_{\mathrm{min}})$ in \eqref{auxiliary_potential}, where $-\lambda_0(r_{\mathrm{min}})$ is the minimum of $-\lambda_0(r)$. Denote by $\mathscr{j}_{l+1/2,1}$ the first positive zero of the Bessel function $J_{l+1/2}$ (see \cite[section 9.5]{AS}). In Proposition~\ref{proposition_non_Efimov_eig} below we give a sufficient condition ensuring that the ground state of \eqref{reduced-radial-higher-momentum} coincides with $E_1^{\text{eff}}$ in \eqref{universal_Efimov_eig}. The proof is presented in Appendix~\ref{App_non_Efimov}

\begin{proposition}\label{proposition_non_Efimov_eig}
    For a fixed angular momentum $l \geq 0$, the ground state of the Hamiltonian \eqref{reduced-radial-higher-momentum} is the first Efimov eigenvalue, if the mass ratio $\frac{M}{m}$ satisfies: 
    \begin{equation}\label{non-Efimov_mass_ratio_condition}
        \frac{M}{m} < \left(\frac{1}{\lambda_0(r_{\mathrm{min}})} \frac{2}{3 \pi}\mathscr{j}_{l+1/2,1} \right)^2 - \frac{1}{8} \, .
    \end{equation}
\end{proposition}

To illustrate \eqref{non-Efimov_mass_ratio_condition}, for $l=0$ (resp. $l=1$) the condition becomes $\frac{M}{m} < 11.24$ (resp. $\frac{M}{m} < 23.12$), to two-decimal accuracy, ensuring that no non-Efimov eigenvalue exists.
\subsection{Spectrum of the effective Hamiltonian away from unitarity}
\begin{figure}[b]
\centering
\includegraphics[width=9cm]{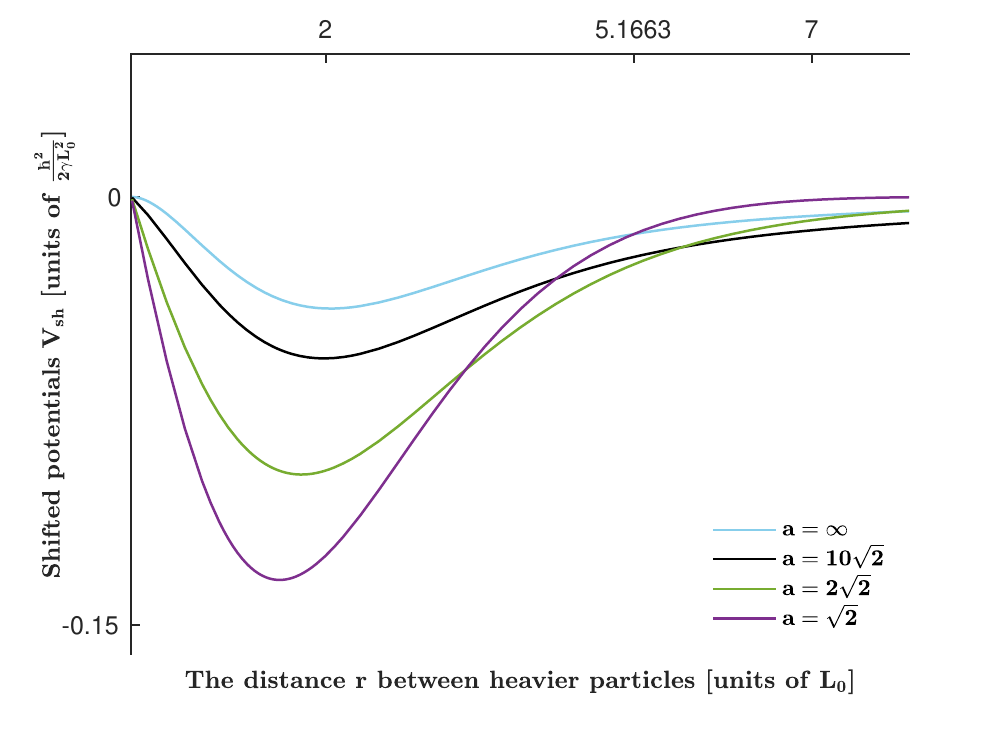}
\caption{\footnotesize{Effective potentials shifted by the constant $(4 \pi \alpha)^2$. The light blue line represents the potential with infinite two-body scattering length $a$. Decreasing $a$ increases the depth of the short-range well.}}
\label{Shifted_potentials}
\end{figure}

In this part we study the eigenvalue problem \eqref{slow} with $V_{\mathrm{eff}}=-\lambda_\alpha$ for $\alpha < 0$. We shift the entire spectrum by the constant $(4 \pi \alpha)^2$ and define
\begin{equation}\label{shifted_lambda}
    V_{\mathrm{sh}}(r,\alpha) := -\lambda_\alpha (r) +(4 \pi \alpha)^2, \qquad \alpha \leq 0 \, .
\end{equation}
In Fig.~\ref{Shifted_potentials} we plot $V_{\mathrm{sh}}$ for $a = \infty,\,10\sqrt{2},\,2\sqrt{2},\,\sqrt{2}$. We immediately observe that, as the two-body scattering length $a$ decreases, the short-range well deepens; consequently, for sufficiently small $a$ the system develops a ground state localized near the origin, essentially independent of the mass ratio.
 
It is natural to expect that, for large $a$, sufficiently close to unitarity, the point spectrum of \eqref{reduced-radial-higher-momentum} consists of low-energy eigenvalues with large spatial extent. Let $\sss$ denote the spatial size of the weakest Efimov-like eigenstate (see Remark~\ref{Efimov-like_eig}). More precisely, $\sss$ should be understood as the maximal spatial size of such a state; following the convention in the physics literature, we simply refer to it as the spatial size. In the Efimov physics literature it is often suggested that $\sss=a$, where $a$ is the two-body scattering length (see, \emph{e.g.}, \cite{Efimov2,Fonseca1,Review}). The underlying idea is to view the three-body system as a superposition of pairwise two-body channels, so that, in the Born--Oppenheimer picture, the effective Hamiltonian \eqref{reduced-radial-higher-momentum} would support bound states localized inside a sphere of radius of order $a$.
 
However, taking $\sss=a$ is not consistent with the actual long-distance behavior of the effective potential. To analyze this more quantitatively, and in analogy with \eqref{k_potentials}, for $\alpha < 0$ (recall that $4 \pi \alpha = -\frac{1}{a}$), we introduce the piecewise potential
 \begin{equation}\label{theta_potentials}
    V_a(r) = \left\{ \begin{aligned}
          -&\lambda_\alpha (r) + (4 \pi \alpha)^2   &\text{if} \quad & 0\leq r \leq a\\
         -&\tfrac{\left(W\left(\ee^{ 4 \pi \alpha r}\right) -4 \pi \alpha r\right)^2}{r^2} + (4 \pi \alpha)^2 &\text{if} \quad & r>a
        \end{aligned}  \right. \, .
\end{equation}
Moreover, for $r>a$, we obtain from Lemma~\ref{analyticity_lemma} the estimate
\begin{equation} \label{long-range_v_theta_estimate}
    V_a(r>a) = 
    -\left(\tfrac{\ee^{-2r/a}}{r^2} + \tfrac{2}{a}\tfrac{\ee^{-r/a} - \ee^{-2r/a}}{r}  \right) + O\left(\tfrac{\ee^{-3r/a}}{r}\right) .
\end{equation}
Up to the term $\frac{2}{a}\frac{\ee^{-2r/a}}{r}$, the long-range behavior \eqref{long-range_v_theta_estimate} coincides with the expressions given in \cite[Eq.~(27)]{Efimov2} and \cite[Eq.~(45)]{Fonseca1}. 
 
In the region that the potential $V_a$ in \eqref{theta_potentials} is below $-\lambda_0(r)$ with inverse-squared behavior, the min–max principle implies that the number of eigenvalues of \eqref{slow} with $V_{\mathrm{eff}}=V_a, \; \alpha <0$ is greater than or equal to that obtained with $V_{\mathrm{eff}}=-\lambda_0(r)$. Therefore, the radius at which these two potentials intersect gives a natural estimate $\sss$.
\begin{corollary}\label{corollary_spatial_size}
    For fixed $ \alpha < 0$, the parameter $\sss$ is given by the first root (other than $r=0$) of $V_a(r)+\lambda_0(r)=0$. In particular, using \eqref{long-range_v_theta_estimate} and \eqref{E02}, one can estimate $\sss$ from
     \begin{equation}\label{spatial_size}
          -\left(\tfrac{\ee^{-2 \sss/a}}{\sss^2} + \tfrac{2}{a}\tfrac{\ee^{-\sss/a} - \ee^{-2\sss/a}}{\sss}  \right) + O\left(\tfrac{\ee^{-3\sss/a}}{r}\right) = -\tfrac{W^2(1)}{\sss^2} \, .
     \end{equation} 
     Therefore, $\sss \simeq 2.80\,a$ to two digits.
\end{corollary}
This means the estimate $\sss \simeq 2.80\,a$ cannot be reconciled with a Hamiltonian built from local point interactions (i.e., constant Bethe–Peierls boundary conditions) in which the three-body system is treated as a sum of pairwise two-body channels. However, this mismatch is not the only issue with that model.
\\
Efimov \cite{Efimov2} and later Fonseca \emph{et al.} \cite{Fonseca1} describe the effective potential in \eqref{slow} away from the unitary limit as having three ranges: a short-range part of size $b$, an intermediate regime with inverse-squared behavior $-\frac{W^2(1)}{r^2}$ for $b < r < a$, and a long-range Yukawa tail for $r\geq a$. In this picture, the intermediate region supports the first $m$ Efimov levels $E^{\mathrm{eff}}_i$, while the remainder of the Efimov spectrum fades into the Yukawa tail. However, this qualitative explanation means that the potential discontinuous at $r=a$ (see Corollary~\ref{corollary_spatial_size} above), which is not physically consistent, and it is in contrary with the actual potential introduced by Fonseca \emph{et al.}.
\\
Furthermore, for any angular momentum $l\ge 0$, Fonseca \emph{et al.} \cite{Fonseca1} applied Bargmann’s bound (see \eqref{Barg_def} below) to the problem \eqref{reduced-radial-higher-momentum} with a potential of form \eqref{theta_potentials} and obtain an upper bound $n_l(V_a)$ on the number of bound states. It is straightforward to see $n_l(V_a)\simeq 0.53\,\frac{\ve^{-2}}{2l+1}$ over the range $a<r<\sss$, independent of the value of $a$. This is rather a large number for the mass-imbalanced system of two heavy and one light particle. For example, in the $l=1$ channel, inserting the critical mass ratio $\mathcal{M}_{1}^{*}$ from \eqref{critical_mass} into $0.53\,\frac{\ve^{-2}}{2l+1}$ gives $n_l(V_a) > 1.16$ on the interval $(a,\sss)$, which is incompatible with the description proposed in \cite{Fonseca1}.
 
Due to the low-energy nature of Efimov physics, Bargmann’s bound (see \cite[Theorem XIII.9]{ReSi}) gives a nearly sharp upper bound on the number of negative eigenvalues of \eqref{reduced-radial-higher-momentum} with $V_\text{eff}=V_a$ in \eqref{theta_potentials}. For fixed $l\ge 0$, let $n_l(V_a)$ denote the number of negative binding energies. Bargmann’s bound is then given by the right-hand side (RHS) of \eqref{Barg_def}:
\begin{equation}\label{Barg_def}
    n_l(V_a) \leq \frac{\ve^{-2}}{(2l+1)} \int_0^{\infty} r \left|V_a(r)\right| \dd r .
\end{equation}
To evaluate the RHS of \eqref{Barg_def}, we split the positive half-line into three intervals: \([0,a]\), \((a, \sss]\), and \((\sss, +\infty)\). On \([0,a]\) we use numerics. For example, for $a=10\sqrt{2}$,
\[
\int_0^{10\sqrt{2}} r \left|V_{10\sqrt{2}}(r)\right|\dd r \simeq 1.1042,
\]
as computed in \textsc{matlab} to four-decimal accuracy.
 
For $r>a$ we use the estimate \eqref{long-range_v_theta_estimate}. Set \(\zeta := r/a\); then, on \((a, \sss]\),
\begin{equation}\label{log-growth}
    \begin{split}
        &\frac{\ve^{-2}}{(2l+1)} \int_{a}^{\sss} \frac{\ee^{-2r/a}}{r} + \frac{2}{a}\left(\ee^{-r/a} - \ee^{-2r/a}\right) \dd r  \\
        =&\frac{\ve^{-2}}{(2l+1)} \left(\int_{1}^{\sss /a} 2\left(\ee^{-\zeta} - \ee^{-2\zeta}\right) + \frac{\ee^{-2\zeta}}{\zeta} \dd\zeta \right)\\
       =& \frac{\ve^{-2}}{(2l+1)} \big(2\ee^{-x}-\ee^{-2x} - \Gamma(0,2x) \big)\Big|_{x=1}^{x=\sss/a} \\ \simeq& 0.5308\, \frac{\ve^{-2}}{(2l+1)} \, ,
    \end{split}
\end{equation}
where we used $\sss = 2.8\, a$ from Corollary~\ref{corollary_spatial_size} and $\Gamma$ is the incomplete gamma function (see \cite[§6.5]{AS}). Similarly, on the interval \((\sss, +\infty)\),
\begin{equation}
        \begin{split}
            &\frac{\ve^{-2}}{(2l+1)} \int_{\sss}^{\infty} \frac{\ee^{-2r/a}}{r} + \frac{2}{a}\left(\ee^{-r/a} - \ee^{-2r/a}\right) \dd r \\  &\simeq 0.1184 \, \frac{\ve^{-2}}{(2l+1)} \, .
        \end{split}
\end{equation}
Note that the growth of the number of bound states proportional to $-\ln a$ is encoded in the incomplete gamma term $\Gamma(0,2x)$. 
 
We conclude this section with a remark on how the energy eigenvalues away from unitarity deviate from the Efimov spectrum in the unitary limit.
\begin{remark}\label{Efimov-like_eig}
    As discussed above, for $\alpha < 0$ the potential $-\lambda_\alpha (r)$ does not exhibit an inverse-squared behavior in any range. Hence, interpreting the eigenvalues that localize within the sphere of radius $\sss$ as genuine Efimov states obeying the geometric law \eqref{universal_geometrical_law} is not accurate. In this regime, the shifted effective potential $V_\text{sh}(r)$ lies below \(-\lambda_0(r)\) for $r < \sss$. Consequently, the wave function associated with any eigenvalue of \eqref{slow} in this case is no longer of the form of modified Bessel function, \emph{i.e.}, $ D\, \frac{ K_{i\beta} (\sqrt{\ve^{-2}\left|E_n\right|}r)}{\sqrt{r}}$ in \eqref{wave_func_Efimov}. This discrepancy grows as \(\alpha\) moves away from unitarity, and for sufficiently large mass ratio this mismatch can even generate additional bound states. For these reasons we refer to these near-threshold states at finite scattering length as \enquote{Efimov-like} eigenvalues, to distinguish them from the Efimov spectrum \eqref{universal_Efimov_eig} that obeys the geometric law \eqref{universal_geometrical_law}.
\end{remark}
\section{Numerical results of parameters of the Efimov spectrum in alkali mixtures and mixture of alkali atoms and a different material}\label{secV}
In this section we compute the binding wavenumbers $\kappa^{(n)}_*$ of the first four Efimov levels in units of $L_0^{-1}$, together with the parameter $\beta$ appearing in \eqref{universal_Efimov_eig} and \eqref{BO_geometrical_law}, for several mixtures of alkali atoms that may be accessible experimentally. The purpose of these computations is mainly illustrative: they show how to apply the model developed in this paper to concrete mass ratios.

In realistic systems, depending on the specific mixture, additional heavy–heavy interactions may need to be included in the effective Hamiltonian \eqref{slow}, for example Lennard–Jones, van der Waals, or dipolar potentials (see \emph{e.g.}, \cite{WangWang2012,PhysRevA.110.033305,z5rx-51yx}). In fact, these additional parameters may help to determine the characteristic length $L_0$.

We first compare our values of $\beta$ with the hyperspherical-coordinate results for $s_0$ reported in \cite{WangWang2012} in the $l=0$ channel of \eqref{reduced-radial-higher-momentum}. We then apply the same framework to $\mathrm{Er}-\mathrm{Li}$ and $\mathrm{Dy}-\mathrm{Li}$ mixtures, which are naturally associated with the $l=1$ channel in \eqref{reduced-radial-higher-momentum}.

Recall that the eigenvalue formula \eqref{universal_Efimov_eig} involves error terms $\zeta_n$, which can be non-negligible for the first few Efimov levels. As discussed after Corollary~\ref{corollary_Efimov_geometrical_law}, these corrections have two sources: (i) the short-range part of the effective potential (see Lemma~\ref{lemma_Ulam}), and (ii) the fact that the inverse-square tail controls the spectrum only asymptotically (see Proposition~\ref{proposition_Efimov_constant_potential}). Consequently, one should expect the ratio in \eqref{BO_geometrical_law} to deviate from an exact geometric value for small $n$, even though $\displaystyle \lim_{n \to \infty} \zeta_n \to 0$.
 
Here we neglect the contribution of $-\lambda_0(r)$ for $r <r_0$ and its exponentially decaying oscillations for $r>r_0$, and approximate $-\lambda_0(r)$ by the auxiliary potential
\begin{equation}  \label{auxiliary_potential_zero}
    V_{\mathrm{aux}}^0(r) = \left\{ \begin{aligned}
          & 0               && \text{if } 0\le r \le r_0\\
         -&\dfrac{W^2(1)}{r^2}   && \text{if } r>r_0
        \end{aligned}\right. 
        \, .
\end{equation}
We compute the next-order correction to \eqref{eqtr1} in the case $\Lambda=0$, \emph{i.e.}, $f(\xi) = \frac{1}{2l+1}$. Note that in the Born–Oppenheimer picture the effective potential has an inverse-square tail $-\frac{W^2(1)}{r^2}$; however, the argument of Lemma~\ref{Lemma_second_order_error} below also applies to hyperspherical-coordinate models, where the tail has the same inverse-square form with a different coefficient. The proof is given in Appendix~\ref{App_Lemma_second_order}.
\begin{lemma}\label{Lemma_second_order_error}
Denote
\begin{equation*}
    -(\eta_n^{0})^2 :=-\frac{4}{ r_0^2} \, \ee^{ \, \frac{2}{\beta} \left(   \arctan \left(\frac{2\beta }{2l+1}\right) + \phi_{\beta,0} -n \pi  \right)} \,
\end{equation*}
and set $\tau_n^{0} := r_0\eta_n^{0}$. In the setting of Proposition~\ref{proposition_Efimov_constant_potential} with $\Lambda=0$, we have:
    \begin{equation*}
        E^{\mathrm{eff}}_n = - \ve^{2}\frac{4}{ r_0^2} \, \ee^{ \, \frac{2}{\beta} \left(   \arctan \left(\frac{2\beta }{2l+1}\right) + \phi_{\beta,0} -n \pi  + \delta_n \right)}
    \end{equation*}
    where 
    \begin{equation}\label{second_order_delta}
        \delta_n=-\frac{\beta\left(1-2 \beta^2\right)}{2\left(1+\beta^2\right)\left((2 l+1)^2+4 \beta^2\right)}\left(\tau_n^{0}\right)^2+O \left(\tau_n^{0}\right)^4 .
    \end{equation}
\end{lemma}
\begin{corollary}\label{Corollary_second_order}
    Assuming $r_0=\sqrt{2}\,\frac{3\pi}{4} L_0$, and $\kappa^{(n)}_* = \sqrt{\ve^{-2} |E^{\mathrm{eff}}_n|}$ we have
    \begin{equation}\label{wave_number_correction}
        \kappa_*^{(n)}=\frac{2}{r_0} \ee^{\frac{1}{\beta}\left(\arctan \frac{2 \beta}{2 l+1}+\phi_{\beta, 0}-n \pi+\delta_n\right)}\, ,
    \end{equation}
    where $\delta_n$ is defined in \eqref{second_order_delta}.
\end{corollary}
From \eqref{wave_number_correction} we see that the correction enters through the factor $\ee^{\delta_n/\beta}$. Moreover, increasing the heavy–light mass ratio increases the size of this correction. Setting 
\begin{equation*}
    \kappa^{(n)}_* = \kappa^{(n)}_0 \ee^{\delta_n/\beta} \, ,
\end{equation*}
we find, for example, that in the mixture $^{174}\mathrm{Yb}_2-^{6}\mathrm{Li}$, the first level $\kappa^{(1)}_*$ differs from $\kappa^{(1)}_0$ by $3.49 \%$, while in $^{133}\mathrm{Cs}_2-^{6}\mathrm{Li}$ the deviation is $2.55\%$. For higher Efimov levels this effect is negligible.

We now briefly comment the calculation $s_0$ in hyperspherical coordinates. denoting $\nu := \arcsin{\frac{M}{M+m}}$, in the $l=0$ channel, the parameter $s_0$ in \eqref{universal_geometrical_law} satisfies a transcendental equation (see \emph{e.g.}, \cite[Eq. (6.8)]{Review})
\begin{equation}\label{transcendental_S-wave_s_0}
    s_0 \cosh{\frac{\pi s_0}{2}} = \frac{2 \sinh (s_0 \nu)}{\sin 2 \nu} \, .
\end{equation}
In the $l=1$ channel, the corresponding equation reads (see, e.g., \cite[Sec.~6.2.2]{Review})
\begin{equation}\label{transcendental_P-wave_s_0}
\frac{1-s_0^2}{s_0} \tan \left(\frac{s_0 \pi}{2}\right)-\frac{2 \cos \left(s_0 \gamma\right)}{\sin 2 \gamma \cos \left(\frac{s_0 \pi}{2}\right)}=\frac{\sin \left(s_0 \gamma\right) }{s_0\sin ^2 \gamma \cos \left(\frac{s_0 \pi}{2}\right)} .
\end{equation}
In Table~\ref{table:alkali_mixtures} we compare $\beta$ and $s_0$ for several mixtures in the $l=0$ and $l=1$ channels, and we also report the dimensionless parameter $\kappa_*^{(n)}L_0$ for $n=1,2,3,4$. Atomic masses are taken from \cite{ProhaskaIrrgeherBenefieldB,Wang_2012_Chemistry,BerglundWieser+2011+397+410}.
\begin{table*}[t]
\caption{Efimov scaling parameter $s_0$, the parameter $\beta$, and the values of $\kappa_*^{(n)}L_0$ for different heavy-heavy-light mixtures.}
\label{table:alkali_mixtures}
\begin{ruledtabular}
\begin{tabular}{lccccccc}
\mixhead & $l$ & $s_0$ & $\beta$ & $\kappa_*^{(1)}L_0$ & $\kappa_*^{(2)}L_0$ & $\kappa_*^{(3)}L_0$ & $\kappa_*^{(4)}L_0$ \\
\hline
\mix{^{174}\mathrm{Yb}_2}{^{6}\mathrm{Li}}   & 0 & 2.246 & 2.117 & 0.294 & 0.064 & 0.015 & 0.003 \\
\mix{^{133}\mathrm{Cs}_2}{^{6}\mathrm{Li}}   & 0 & 1.983 & 1.840 & 0.229 & 0.041 & 0.007 & $1.33\times 10^{-3}$ \\
\mix{^{87}\mathrm{Rb}_2}{^{6}\mathrm{Li}}    & 0 & 1.632 & 1.468 & 0.148 & 0.017 & 0.002 & $2.37\times 10^{-4}$ \\
\mix{^{41}\mathrm{K}_2}{^{6}\mathrm{Li}}     & 0 & 1.154 & 0.962 & 0.052 & 0.002 & $7.57\times 10^{-5}$ & $2.89\times 10^{-6}$ \\
\mix{^{23}\mathrm{Na}_2}{^{6}\mathrm{Li}}    & 0 & 0.875 & 0.667 & 0.014 & $1.27\times 10^{-4}$ & $1.15\times 10^{-6}$ & $1.03\times 10^{-8}$ \\
\mix{^{87}\mathrm{Rb}_2}{^{40}\mathrm{K}}    & 0 & 0.653 & 0.424 & $1.15\times 10^{-3}$ & $7.04\times 10^{-7}$ & $4.30\times 10^{-10}$ & $2.62\times 10^{-13}$ \\
\mix{^{133}\mathrm{Cs}_2}{^{87}\mathrm{Rb}}  & 0 & 0.536 & 0.276 & $2.49\times 10^{-5}$ & $2.87\times 10^{-10}$ & $3.32\times 10^{-15}$ & $3.83\times 10^{-20}$ \\
\mix{^{167}\mathrm{Er}_2}{^{6}\mathrm{Li}}   & 1 & 2.102 & 1.514 & 0.115 & $1.44\times 10^{-2}$ & $1.81\times 10^{-3}$ & $2.27\times 10^{-4}$ \\
\mix{^{166}\mathrm{Er}_2}{^{6}\mathrm{Li}}   & 1 & 2.095 & 1.506 & 0.113 & $1.40\times 10^{-2}$ & $1.74\times 10^{-3}$ & $2.16\times 10^{-4}$ \\
\mix{^{164}\mathrm{Dy}_2}{^{6}\mathrm{Li}}   & 1 & 2.083 & 1.488 & 0.110 & $1.33\times 10^{-2}$ & $1.61\times 10^{-3}$ & $1.94\times 10^{-4}$ \\
\mix{^{163}\mathrm{Dy}_2}{^{6}\mathrm{Li}}   & 1 & 2.077 & 1.479 & 0.108 & $1.30\times 10^{-2}$ & $1.54\times 10^{-3}$ & $1.84\times 10^{-4}$ \\
\mix{^{161}\mathrm{Dy}_2}{^{6}\mathrm{Li}}   & 1 & 2.064 & 1.460 & 0.105 & $1.21\times 10^{-2}$ & $1.42\times 10^{-3}$ & $1.65\times 10^{-4}$ 
\end{tabular}
\end{ruledtabular}
\end{table*}
The quantities reported in Table~\ref{table:alkali_mixtures} are dimensionless and the physical wavenumbers are obtained only after specifying the characteristic length  $L_0$. This suggests also that the parameter $L_0$ may depend on the mixture. In particular, a direct comparison with experimentally measured Efimov features for $\mathrm{Cs}-\mathrm{Li}$ mixture in \cite{PhysRevLett.113.240402,PhysRevLett.112.250404} cannot by itself determine \(L_0\). Such a comparison is affected by the difference between the Born--Oppenheimer parameter \(\beta\) and the hyperspherical parameter \(s_0\), and also theoretical complexity of the nature of the approximation in the Born--Oppenheimer picture (see Sec~\ref{secVI} below). Moreover, additional heavy--heavy interactions are present in realistic mixtures. Nevertheless, the comparison may still be useful, since it can indicate the order of magnitude of \(L_0\), for example a van der Waals length or another short-distance length. Additionally, if the minimizer of $V_{\mathrm{eff}}$ is denoted by $r_{\mathrm{min}}$, then, for sufficiently large mass ratios, which are more relevant to molecular physics, one can apply the Born--Oppenheimer scheme and expect the ground state of the system to have spatial size approximately $r_{\mathrm{min}}$ (see the discussion in Sec.~\ref{secVI}, and in particular the approximation in \eqref{expansion_BO_mol}). This fact also helps us to determine $L_0$. However, a precise identification of \(L_0\) (if possible) requires further analysis and is left for future work.
\section{Conclusions and perspectives}\label{secVI}
In this work, we studied a $2+1$ mass-imbalanced system in the Born--Oppenheimer approximation, where the heavy–light interactions were modeled by zero-range (delta-type) interactions.
Using von Neumann’s theory, we defined a family of light-particle Hamiltonians which, unlike local point (delta) interactions (constant Bethe–Peierls conditions \eqref{BP_BC_two-body}), remain regular as the distance between the heavy centers tends to zero. The effect of these two-body zero-range interactions resembles a three-body force in the sense that the distance between the scattering centers enters explicitly into the two-body boundary conditions (see \eqref{3D_bos_Bethe_BC} and \eqref{3D_fer_Bethe_BC}).

We showed that this construction is consistent with the expected low-energy behavior of a heavy–heavy–light system. However, unlike short-range potential models, this model does not provide the physical length scale and it must be supplied externally. After fixing this characteristic length, in the unitary limit we derived an analytic formula for the Efimov spectrum. We also provided a sufficient condition ensuring that the ground state coincides with the first Efimov level. Moreover, we illustrate the method by computing the Efimov scaling parameter $\beta$ and the binding wavenumbers of the first four Efimov levels in units of the inverse characteristic length, for several mixtures.

We also stress that the short-distance prescription used in this work is not unique. This should not be viewed as a drawback of the construction, but rather as a standard feature of contact three-body theories. Zero-range models are effective descriptions of the long-distance regime and are not expected to reproduce the microscopic interaction at arbitrarily small distances. Therefore, the role of the regularization is to remove the Thomas collapse and to fix the required three-body input.

Away from unitarity, we followed the deformation of the spectrum and showed how qualitative pictures based on local point interactions conflict with the quantitative features of the three-body problem. In particular, we found that the size of the weakest near-threshold trimer is approximately three times larger than the two-body scattering length, in contrast to the common assumption that these two length scales coincide. Additionally, we refined Bargmann’s bound on the number of bound states and analyzed how the spectrum away from the unitary limit differs from the Efimov spectrum in the unitary regime. Finally, we comment on the validity of the Born--Oppenheimer approximation for Efimov physics.

The validity of the Born–Oppenheimer approximation in Efimov physics is often taken for granted once the mass ratio exceeds the critical value \eqref{critical_mass}. However, the comparison between $\beta$ and $s_0$ in Table~\ref{table:alkali_mixtures} indicates that this expectation is not quantitatively justified. Denoting by $E_n^{\mathrm{2+1}}$ the Efimov energies of the full $2+1$ Hamiltonian $\hh^{b/f}_\ve$, for $k>0$, one would expect to have an asymptotic expansion of the form
\begin{equation}\label{expansion_BO_Efimov}
    E_n^{\mathrm{2+1}} = E_n^{\mathrm{eff}} + O(\ve^k)\, .
\end{equation}
The Ansatz \eqref{expansion_BO_Efimov} is inspired by the standard Born–Oppenheimer theory in molecular physics, where the approximation is primarily used to describe states localized near the minimum of a smooth effective potential. Let $E_0$ denote the corresponding short-range ground-state energy (to distinguish it from the Efimov spectrum). In this picture, $V_{\mathrm{eff}}(r_{\mathrm{min}})$ is a lower bound for $E_0$, where $r_{\mathrm{min}}$ is the minimizer of $V_{\mathrm{eff}}$. Here, the kinetic energy term $-\ve^2 \Delta_{\R}$ raises the ground-state energy slightly above the value $V_{\mathrm{eff}}(r_{\mathrm{min}})$. Moreover, for any fixed $n$, there exists a mass parameter $\ve>0$ sufficiently small such that the system has (at least) $n+1$ isolated eigenvalues of the form
\begin{equation}\label{expansion_BO_mol}
    E^n = E_0 + \ve \, C_n E_0+ O(\ve^2) \,
\end{equation}
where the coefficients $C_n$ are obtained from a harmonic approximation and depend on the shape of $V_{\mathrm{eff}}(r)$ in a neighborhood of its minimum \cite{Simon1983}. Such expansions can change in the presence of nonsmooth minima: for instance, if the effective potential has a sharp twist at its minimum, the leading correction becomes of order $\ve^{2/3}$, with coefficients expressed in terms of zeros or extrema of the Airy function (\cite{Duchene_2014, cacciapuoti2025born}).

In the Efimov setting, the validity of an expansion like \eqref{expansion_BO_Efimov} is far less clear. In particular, the hyperspherical exponent $s_0$ obtained from \eqref{transcendental_S-wave_s_0} and \eqref{transcendental_P-wave_s_0}, depends on the mass ratio and satisfies $s_0 \to \infty$ as $M \to \infty$. Thus, while $\beta$ approaches $s_0$ in the limit $\ve \to 0$ (equivalently $M \to \infty$), long-range shallow eigenvalues collapse towards the threshold in this regime.

There is also a natural difference with the molecular situation. The expansion \eqref{expansion_BO_mol} describes a ladder of states increasingly concentrated near the potential minimum as $\ve \to 0$, so the Born–Oppenheimer approximation becomes more accurate for the low-lying spectrum. In Efimov physics, by contrast, the states of interest are long-range excited trimers whose energies are not governed by the structure of $V_{\mathrm{eff}}$ near its minimum. Heuristically, as $\ve \to 0$ the minimum of $V_{\mathrm{eff}}$ controls the ground state and produces many nearby levels separated on an $\ve$-scale, while the long-range Efimov states fade away towards threshold.

Finally, typical mass ratios in molecular systems can be extremely large, \emph{i.e.}, $1.8\times 10^{3}$–$3.7\times 10^{5}$, well beyond those encountered in ultracold atomic mixtures where Efimov physics is studied. Taken together, these observations lead us to question the Born--Oppenheimer approximation as a quantitative method for determining three-body eigenvalues in Efimov physics. 

At the same time, the appearance of the Efimov effect in this framework suggests that there is still a meaningful connection. To clarify this, we briefly recall how one may attempt to relate a $2+1$ Hamiltonian $\hh^{b/f}_\ve$ to the light-particle Hamiltonian $\hh^{b/f}_\alpha$.

In the literature (see, e.g., \cite{Krejcirik-etal-m2018}), it is common to introduce the direct integral of the light-particle Hamiltonian (more precisely, of its quadratic form and the associated operator),
\begin{equation*}
    \hh^{b/f}_{\alpha,\x} := \int_{\rr^3} ^{\oplus} \hh^{b/f}_\alpha (\R) \, \dd \R \, ,
\end{equation*}
and (formally) rewrite the $2+1$ operator $\hh^{b/f}_\ve$ as
\begin{equation}\label{direct_sum}
    \hh_\ve^{b/f}=  -\ve^2 \Delta_{\R} + \hh^{b/f}_{\alpha,\x} \, .
\end{equation}
For smooth heavy–light interactions such a decomposition can be justified. In three (and two) dimensions with zero-range interactions, however, \eqref{direct_sum} is not valid. Roughly speaking, \eqref{direct_sum} enforces a strict separation between the $\x$- and $\R$-dependence, while the zero-range coupling is implemented through TMS contact conditions on the coincidence hyperplanes $\x = \pm \R/2$. These conditions are intrinsically \enquote{radial} in the configuration space and cannot be encoded by simply taking a direct integral over $\R$. Moreover, passing to hyperspherical coordinates does not resolve this issue. Rather, at the level of quadratic forms, the form associated with $\hh_\ve^{b/f}$ contains cross terms that do not arise from the direct integral construction (see, e.g., \cite{Basti2021ThreeBodyHW,10.1007/978-981-99-5894-8_8} for three identical bosons; the same strategy applies to the $2+1$ problem). Consequently, the natural identification $ \theta_{\mathrm{3b}}(r) = g_\alpha(r)$ (or any other admissible renormalization function $\theta(r)$; see Remark~\ref{remark_theta_function}) does not, by itself, recover the renormalized TMS contact conditions \eqref{TMS_bosonic_BC} and \eqref{TMS_fermionic_BC} from the Bethe–Peierls–type boundary conditions \eqref{3D_bos_Bethe_BC} and \eqref{3D_fer_Bethe_BC}.

It is also worth noting that even in one spatial dimension, for the formal operator
\begin{equation*}
    -\ve^2\partial^2_R -\partial^2_x + \alpha \delta(x \pm R/2), \quad x,R \in \rr \,,
\end{equation*}
the identity \eqref{direct_sum} still fails at the operator level. In particular, a factorized ansatz $\Psi(x,R) = \psi(x,R) \Phi(R)$ does not belong to the domain of the $2+1$ Hamiltonian, because it fails to satisfy the required boundary conditions in the $R$-variable. This obstruction can nevertheless be bypassed at the quadratic-form level; see \cite{cacciapuoti2025born} for details.

This viewpoint suggests that, in Efimov physics, understanding the accuracy of the Born–Oppenheimer approximation is closely tied to the problem of constructing a genuine three-body Hamiltonian starting from the light-particle model. In a sense, this is the reverse direction compared with the usual molecular-physics setting.

To summarize, the quantitative validity of the Born--Oppenheimer approximation in Efimov physics is far from understood. Nevertheless, this formal framework captures the qualitative features of the Efimov effect and provides a better understanding of the three-body problem, whose spectral analysis is considerably harder and more involved.


\appendix
\section{Construction of one- and two-Center point-interaction Hamiltonians via von Neumann’s theory}\label{App_construction}
This appendix is intended for readers who may not be familiar with self-adjoint extensions of densely defined symmetric operators. For further background and details, we refer to \cite[Appendix~A]{Albeverio} and to \cite{DG,reed_methods_1975}.

To build intuition, we first recall von Neumann’s construction in the standard one-center case. Formally, one writes the Hamiltonian as
\begin{equation*}
    A_\al:= -\Delta_{\x} - \mu \delta(\x-\y) \, ,
\end{equation*}
where $\y \in \rr^3$ denotes the position of a fixed scattering center. It is well known that functions $\psi$ in the domain $D(A_\al)$ decompose into a regular part $u \in H^2(\rr^3)$ (the domain of the free Laplacian) plus a singular contribution that enforces the Bethe–Peierls behavior \eqref{BP_BC_two-body} near $\y$. In what follows, we show that the operator $A_{\phi}$ obtained rigorously via von Neumann’s extension theory is equivalent to the Hamiltonian $A_\al$.
 
The Hamiltonian $A_\al$ is rigorously defined as the perturbation of the free Laplacian, supported on the point $\y$: 
\begin{equation*}
    \dot{A}_{0} = -\Delta_{\x} \restriction C^{\infty}_{0}(\mathbb{R}^3 \setminus \{\y\}) \, .
\end{equation*}
The operator $\dot{A}_{0}$ is densely-defined and unbounded on $L^2(\rr^3)$, and it is symmetric. However, it is not self-adjoint: the domain of its adjoint $A^{\dagger}$ is strictly larger than $D(\dot{A}_{0})$. A convenient way to see this is through the Green’s function $G_{z}$ of the Helmholtz operator, defined by 
\begin{equation*}
    \left(-\Delta - z\right) {G}_{z}(\x)  = \delta (\x), \quad z \in \cc \setminus \rr^+ \, .
\end{equation*}
Although $G_z \notin D(\dot{A}_{0})$, for any $\phi \in D(\dot{A}_{0})$, one has
\begin{equation*}
    \langle{G}_{z} , \dot{A}_{0}\phi\rangle = \langle z{G}_{z} , \phi\rangle \,,
\end{equation*}
which shows that $G_z \in D(A^{\dagger})$. Explicitly,
\begin{equation}\label{Green's_function}
    {G}_{z}(\x) = \frac{\ee^{i \sqrt{z} |\x|}}{4 \pi |\x|}, \quad \Im z >0.
\end{equation}
Intuitively, to obtain a self-adjoint extension of the symmetric operator $\dot{A}_{0}$, one enlarges its domain inside \(D(A^\dagger)\) by adding suitable defect vectors, in such a way that the resulting operator is self-adjoint. To describe all such extensions, we introduce the deficiency (defect) spaces of $\dot{A}_{0}$, denoted by $N_{z}$ and $N_{z^*}$. For $z \in \cc \setminus \rr^{+}$, we set
\begin{equation}\label{defect_spaces_def}
    N_{z} := \ker \left(A^{\dagger} - z^*\right), \quad N_{z^*} := \ker \left(A^{\dagger} - z\right)
\end{equation}
 A closed symmetric operator admits self-adjoint extensions if and only if the dimensions of the defect spaces are equal (see \emph{e.g.}, \cite[section X.1, p. 141]{reed_methods_1975}). In the present one-center case, both spaces are one-dimensional, and the Green’s function $G_z$ (resp. $G_{z^*}$) in \eqref{Green's_function} span $N_z$ (resp. $N_{z^*}$). Von Neumann’s theory states that self-adjoint extensions are in one-to-one correspondence with unitary maps
\begin{equation*}
    J : N_{z} \to N_{z^*} \, .
\end{equation*}
Since $\dim N_z =1$, such a unitary operator $J$ identifies by $\ee^{i \phi}$ for $\phi \in [0, 2 \pi)$. Denote by $A_\phi$ the corresponding self-adjoint extension. Then, by von Neumann’s decomposition, any $f \in D(A_\phi)$ can be written as
\begin{equation}\label{von_Neumann_decomposition_domain}
     f =  f_0 +  f^z + Jf^{z},\, \quad f_0 \in D(\dot{A}_{0}), f^z \in N_z \, .
\end{equation}
and the action of $A_\phi$ is given by
\begin{equation}\label{von_Neumann_decomposition_action}
    A_\phi f = \dot{A} f_0 + z f^z + z^* Jf^{z} \, .
\end{equation}
Here, for a general defect parameter $z=i\kappa^2$ the parameter \(\kappa>0\) has physical dimension \([L]^{-1}\). Different choices of $\kappa$ only correspond to different parametrizations of the same family of self-adjoint extensions. We therefore choose the length unit so that $\kappa =1$. In physical variables, this corresponds to measuring lengths in units $1/\kappa$, which we identify with the characteristic length $L_0$.

We see that, von Neumann’s decomposition for $f \in D(A_\phi)$ gives 
\begin{equation*}
    f = f_0 + \frac{\ee^{i \sqrt{i} |\x - \y|}}{4 \pi |\x - \y|} + \ee^{i \phi} \frac{\ee^{i \sqrt{-i} |\x - \y|}} {4 \pi |\x - \y|}, \quad \Im \sqrt{\pm i} >0 .
\end{equation*}
The extension parameter \(\phi\) is related to the point-interaction parameter \(\alpha\) by
\[
4\pi\frac{\alpha}{\kappa}
=
\cos\frac{\pi}{4}
\left(\tan\frac{\phi}{2}-1\right).
\]
Here, $\frac{\al}{\kappa}$ is dimensionless. After fixing the reduced units by setting \(\kappa=1\), we then (by abuse of notation), we keep the same notation \(\alpha\) for the
extension parameter and identify it with the reduced scattering length through
\[
4\pi\alpha=-\frac{1}{a}.
\] 
We recall that $a = a_s/ L_0$, where $a_s$ is the physical scattering length.

Equivalently, for $z=z^*=- \lambda, \,\lambda \in \rr^+$ and denoting $G_\lambda := G_{z=-\lambda}$, the domain of the one-center Hamiltonian $A_\al$ can be written as
\begin{widetext}
    \begin{equation}\label{one-center_domain}
D\left(A_\al\right)=\left\{\psi \in L^2 \mid \psi(\x)=u^\lambda+q G_\lambda(\x-\y), u^\lambda \in H^2, \, q \in \mathbb{C}, \,u^\lambda(\y)=\left(\alpha+\frac{\sqrt{\lambda}}{4 \pi}\right) q\right\}
\end{equation}
\end{widetext}
Here, $q$ is the (one-center) point charge, analogous to the boundary charge $\xi$ in \eqref{TMS_bosonic_BC}–\eqref{TMS_fermionic_BC}. In physical variables, it has dimension $[L]^{-1/2}$. In the reduced length units used here, this length dimension is absorbed into the scale \(L_0\). Setting $\rho := |\x- \y|$, the matching condition in \eqref{one-center_domain} implies the Bethe–Peierls behavior \eqref{BP_BC_two-body}. Indeed, as $\rho \to 0$ one has, up to an $o(1)$ term,
\begin{equation*}
    \psi (\x) = u^\lambda(\y)+q \left(\frac{1}{4 \pi \rho} - \frac{\sqrt{\lambda}}{4\pi}\right) = q \left(\frac{1}{4 \pi \rho} + \al\right) \, ,
\end{equation*}
which is equivalent to \eqref{BP_BC_two-body}.
 
We now turn to the operator $\hh^{b/f}_\al$ that appears formally in \eqref{fast}. As in the one-center case, we construct it as a self-adjoint extension of the symmetric, densely defined operator
\begin{equation*}
    \dot{\hh}_{0} = -\Delta \restriction C^{\infty}_{0}(\mathbb{R}^3 \setminus \{\y_1,\y_2\}) \, .
\end{equation*}
Self-adjoint extensions are parametrized by the action of a unitary map $J$ between the defect spaces. We denote the relevant extension parameter by $\theta$, and later relate $\theta$ to $\alpha$ and derive the resolvent of $\hh^{b/f}_\ve$.
 
For the two-center point-interaction problem, the defect spaces $N_{i}$ and $N_{-i}$ of $\dot{\hh}_0$ are both two-dimensional. A convenient choice of bases is given by the Green’s functions centered at $\y_1$ and $\y_2$:
\begin{equation}\label{Green_i_functions}
G_{\pm i}^{(n)}(\x \, ; \y_n) = \frac{e^{i \sqrt{\pm i}|\x -\y_{n}|}}{4\pi|\x - \y_{n}|}, \quad \, n=1,2, \quad \x \in \rr^3 \,. \nonumber
\end{equation}
Without imposing any symmetry, the defect spaces can be written as
\begin{equation} \label{defect_spaces}
    N_{\pm i} = \left\{ f \in L^2(\rr^3) \,|\, f = c_1 G^{(1)}_{\pm i} + c_2 G^{(2)}_{\pm i}\right\}, \quad c_1,c_2 \in \cc \, .
\end{equation}
A unitary map $J: N_{i} \to N_{-i}$ is therefore represented, in these bases, by a $2 \times 2$ unitary matrix $U$; explicitly,
\begin{equation*} 
 J G_{i}^{m} = \sum_{n=1}^{2}U_{nm} \, G_{-i}^{n}
\end{equation*}
Consequently, for $ f^i= c_1 G^{(1)}_{ i} + c_2 G^{(2)}_{ i}$
\begin{equation*}
    J f^i=J\left(\sum_{m=1}^2 c_m G_i^{(m)}\right)=\sum_{n=1}^2\left(\sum_{m=1}^2 U_{n m} c_m\right) G_{-i}^{(n)} .
\end{equation*}
Furthermore, we assume:
    \begin{itemize}
        \item[-] the scattering centers are identical. This means the matrix $U$ is constrained to be diagonal, with each diagonal element expressed as $e^{i \theta}$, where $0 \leq \theta < 2 \pi$;
        \item[-] the wave function of the light particle under the exchange the positions of the centers, either remains unchanged or flips its sign. The symmetry (resp. anti-symmetry) assumption means $c_1 = c_2$ (resp. $c_1 = -c_2$) in \eqref{defect_spaces}.
    \end{itemize}
Henceforth, for simplicity, we restrict to the symmetric/antisymmetric subspaces by choosing $c_1,c_2 \in \{1,-1\}$. We denote by $\hh^{b}_\theta$ (resp. $\hh^{f}_\theta$) the corresponding self-adjoint extension of $\dot{\hh}_0$ in the symmetric (resp. antisymmetric) case. Due to the fact that a two-center point interaction is a rank-two perturbation of the Laplacian, the extension theory and the resolvent formula naturally involve a $2 \times2$ structure: the unitary map $J$ acts through a $2 \times2$ unitary matrix, and the resolvent of $\hh^{b/f}_\theta$ contains a $2 \times2$ matrix encoding this choice. We now make this relation explicit.

The free Laplacian $-\Delta_\x$ is the \enquote{trivial} self-adjoint extension of $\dot{\hh}_0$. Denote its resolvent by $(-\Delta_{\x} - z)^{-1}$, for $\x, \x' \in \rr^3$, $z \in \cc \setminus \rr^+$ and $f \in L^2(\rr^3)$, we apply the Krein's formula for the difference of resolvents of two self-adjoint extensions:
\begin{widetext}
    \begin{equation}\label{Kerin's_formula}
    \left[(\hh^{b/f}_\theta - z)^{-1}f\right](\x) = \left[(-\Delta_{\x} - z)^{-1}f\right](\x) +   \sum_{m,n =1}^{2}  \bigl[\Gamma^{b/f}_\theta (z)^{-1}\bigr]_{mn}
       G_{z}^m(\x-\y_m) \int_{\rr^3} G_z^n(\y_n - \x') f(\x') \, \dd \x' \, .
\end{equation}
\end{widetext}
Here, the matrix $\Gamma^{b/f}_\theta (-\lambda)^{-1}$ satisfies
\begin{widetext}
    \begin{equation}
    \bigl[\Gamma^{b/f}_\theta (z)^{-1}\bigr] - \bigl[\Gamma^{b/f}_\theta (z_0)^{-1}\bigr] = (z - z_0) \bigl[\Gamma^{b/f}_\theta (z)^{-1}\bigr] S(z^*,z_0) \bigl[\Gamma^{b/f}_\theta (z_0)^{-1}\bigr] \, ,
\end{equation}
\end{widetext}
where, for $1\leq m,n \leq 2$, elements of matrix $S$ are defined by:
\begin{equation}\label{matrix_S_def}
    S_{mn} (z^*{,}w) = \int_{\rr^3} \bigl(c_mG^m_{z}\bigr)^*  c_nG^n_w  \,\dd \x, \quad z,w \in \cc \setminus \rr^+ \,.
\end{equation}
Dąbrowski and Grosse applied the Krein's formula at $z_0= -i$ and connect the unitary matrix $U$ (representing $J$) to $\Gamma^{b/f}_\theta (-\lambda)^{-1}$; see \cite[Eqs.~(2.18)–(2.20)]{DG}:
\begin{equation} \label{Matrix-Gamma}
\Gamma^{b/f}_\theta(z) =  2i S(i,i)(U^{T}+I)^{-1} - (z+i)S(z^*,-i).
\end{equation}
In our symmetric/antisymmetric reduction, $ 2i(U^{T}+I)^{-1}$ forms a diagonal matrix, with each diagonal element being $i + \tan{\frac{\theta}{2}}$. We denote $t_\theta := \tan{\frac{\theta}{2}}$ which ranges over $(-\infty , \infty)$ labels the self-adjoint extensions, and is closely related to the two-body scattering length (see \eqref{identification_von_Neumann_al} below). Moreover, from \eqref{matrix_S_def} we have
\begin{equation}\label{matrix_S_entries}
\begin{aligned}
&S_{mn}(i, i)  =\frac{1}{4 \sqrt{2} \pi}, \, &\text{for}\, \, m=n \\
&S_{mn}(i, i)  =\pm\frac{1}{4 \pi r} \mathrm{e}^{-r / \sqrt{2}} \sin \left(\frac{r}{\sqrt{2}}\right), \, &\text{for}\, \, m\neq n  \\
-(z+i) &S_{mn}(z^*, -i)  =\frac{\sqrt{-z}-\sqrt{i}}{4 \pi}, \, &\text{for}\, \, m=n  \\
-(z+i) &S_{mn}(z^*, -i)  =\pm\frac{\mathrm{e}^{-\sqrt{i} r}-\mathrm{e}^{-\sqrt{-z} r}}{4 \pi r} , \,\, &\text{for}\,  m \neq n 
\end{aligned}
\end{equation}
where \(+\) (resp. \( - \)) in off-diagonal terms corresponds to exchange symmetry $\hh^{b}_\alpha$ (resp. exchange anti–symmetry $\hh^{f}_\alpha$) case (see also \cite[Eq. (16)]{FST} for $n$-center case). Also, we set $z = z^* = -\lambda, \, \lambda \in \rr^+$, and identify  
\begin{equation}\label{identification_von_Neumann_al}
    \alpha :=\frac{t_\theta - 1}{4\sqrt{2}\pi} \, .
\end{equation}
If we put \eqref{matrix_S_entries} in \eqref{Matrix-Gamma}, we obtain \eqref{Gamma-nonlocal}.

\begin{remark}\label{extension_parametrization}
    We emphasize that the choice of the defect parameter $\kappa$ does not change the self-adjoint operator, provided the unitary map between the defect spaces is transformed consistently. In fact, the invariant object is the resolvent itself; it would be just a different parameterization of a same operator. Consequently, if one chooses $z= i \kappa^2$ and $z^* = -i \kappa^2$ with $\kappa \neq 1 $ as the defect parameters, the matrix $U$ that represents the unitary operator $J$ between the defect spaces has a different shape and is not simply of diagonal form.
\end{remark}
\begin{remark}\label{remark_theta_function}
    We remark that, in the construction presented in this appendix, the extension parameter $t_\theta$ is kept fixed as the separation $r$ varies. This is a natural choice, but it is still a special one. Dąbrowski and Grosse characterize all self-adjoint extensions of the symmetric operator $\dot{\hh}_{0}$, for fixed positions of the scattering centers; in particular, the distance $r$ between the centers is fixed in their construction. In the Born--Oppenheimer scheme, however, $r$ is a variable. More generally, one may therefore allow the extension parameter $\theta$, equivalently the function $t_\theta$, to depend on $r$. From this point of view, replacing $g_\alpha$ in the off-diagonal entries of $\Gamma^{b/f}_{\alpha}(-\lambda)$ in \eqref{Gamma-nonlocal} by a suitable function $\theta(r)$ corresponds to choosing a family of self-adjoint extensions, one for each value of $r$. In this sense, the quadratic-form construction can be interpreted, in von Neumann terms, as gluing together different fiber operators as $r$ varies.
    
    Note that the choice of $\theta(r)$ is not arbitrary if one wants the resulting effective Hamiltonian \eqref{slow} to have the desired stability and coincidence-limit properties. In particular, the behavior of $\theta$ near $r=0$ s crucial; if $\theta(0) \neq 1$, the stability of the effective Hamiltonian is not guaranteed (see also the remarks at the end of Sec~\ref{secIII}).
    \end{remark}
\section{Proof of Lemma~\ref{analyticity_lemma}} \label{appendix1}
Define
         \begin{equation} \label{omega_r_theta}
            \omega(r, \alpha) := 4 \pi \alpha r + g_\alpha(r), \qquad r \geq 0 \,.
        \end{equation}
It is evident that \( u = f^{-1} \left(\omega\right) = W(\ee^{\omega}) - \omega\) is the inverse function of \(\omega = f(u) = \ee^{-u} - u\). The main tool here is the Lagrange inversion theorem (see \cite[section 3.6]{AS}), \emph{i.e.}, \(f^{-1}\) is analytic at \(\omega_0  = f(u_0)\), if \(f^{\prime}(u_0)\neq 0\), provided \(f\) is analytic itself. 
    \\
Clearly, $\omega(r, \alpha)$ in \eqref{omega_r_theta} is a composition of analytic functions, so it is analytic itself. To satisfy the conditions of Lagrange inversion theorem, $\omega_0 = f(u_0)$ must belong to a domain $\mathcal{V}$ that
     \begin{equation*}
         f^{\prime}(u_0)\neq 0, \;\; f(u_0) \in \mathcal{V} \, .
     \end{equation*}
Since \(f'(u)=-(\ee^{-u}+1)\), the critical points $u_c=-\,i(2k+1)\pi,\, k\in\mathbb Z$ satisfy \(f'(u)=0\). The corresponding critical values under the map $f$ are
\[
\omega_c=f(u_c)=-1+i(2k+1)\pi,\qquad k\in\mathbb Z.
\]
The ones closest to the real axis are $\omega_c^{\pm}=-1\pm i\pi$. By Lagrange inversion, the Taylor series of \(f^{-1}\) about \(\omega_0\)
converges in the open disk centered at \(\omega_0\) with radius equal to
the distance from \(\omega_0\) to the set of critical values:
\[
\kappa:=\operatorname{dist}\big(\omega_0,\{\omega_c\}_{k\in\mathbb Z}\big)
=\inf_{k\in\mathbb Z}\, \big\lVert\omega_0-(-1+i(2k+1)\pi)\big\rVert.
\]
In particular, if \(\omega_0 \in \rr\) in this case, then the nearest critical values are
\(\omega_c^{\pm}\), so $\kappa=\lVert\omega_0-\omega^\pm_c \rVert $.

For \(\omega_0=\omega(r, \alpha)\) from \eqref{omega_r_theta} with \(\alpha \leq 0 \) and \(r\ge0\), we have \(\kappa>0\), so the series for \(f^{-1}\) is valid. Moreover, the factor \(r^{-2}\) is a removable singularity at \(r=0\);
hence \(-\lambda_\alpha (r)\) extends analytically to the real half-line, and the asymptotics \eqref{E01}–\eqref{E02} follow.
 
At \(r=0\) we have \(\omega(0,\alpha)=1\) for any $\alpha \leq 0$. Hence, $\kappa=\sqrt{\pi^2+4}$ and the Taylor series of \(f^{-1}\) about \(\omega_0=1\) has radius \(\kappa=\sqrt{\pi^2+4}\). However, the distance that we can vary \(r\) while keeping \(\omega(r, \alpha)\) inside the disk \(\mathbb D(1,\kappa)\) depends on \(\alpha\). Set
\[
u(r):=f^{-1}(\omega(r, \alpha)) \, ,
\]
there exists \(\delta(\alpha)\) the range that \(|\omega(r, \alpha)-1|<\kappa\) for \(0\le r<\delta(\alpha)\). We readily see the admissible \(r\)-range shrinks as \(\alpha\to-\infty\), so
\(\delta(\alpha)\to0\). Moreover, at unitarity \(\alpha = 0\) , we have
\(-\sqrt2\,\ee^{-5\pi/4}\le \omega(r,1)\le 1\), so \(\omega(r,1)\in\mathbb D(1,\kappa)\) for all \(r\ge0\). Therefore, the admissible \(r\)-range is unbounded radius to the right in the \(r\)-variable.

Next, we now show that, for \(\alpha < 0\), the Taylor series in \(r\) of $u(r)$
about \(r=a\) converges for all \(r\ge a\) \emph{i.e.}, it is unbounded in the variable $r$ as well. So, at \(r=a\), we have $\omega_0 := \omega\big(a;\alpha\big)=-1+g_\alpha\big(a\big)$. With the same reasoning as before, the Taylor series of \(f^{-1}\) is convergent about $\omega_0$ with the radius $\pi$. Therefore, if $|\omega(r, \alpha)|  \leq \pi$, Taylor series of $u(r)$ about \(r=a\) converges
for every \(r\ge a\). Additionally, if $\omega(r, \alpha)  \leq -1$, the Taylor series of the term $W(\ee^{\omega})$ is convergent for any $r$ and again, the radius of convergence of Taylor series of $u(r)$ about the point $a$ is infinite to the right. This because the principal branch \(W_0(x)\) has a Taylor expansion about \(x=0\) with radius \(\ee^{-1}\) (Corless, \emph{et al.} \cite{Corless:1996zz}).
Summing up these two conditions, we need to show $\omega(r, \alpha)  \leq \pi$ for $r > a$. We verify it in two ranges of \(\alpha\).

1) If \(-(\frac{1}{4 \sqrt{2} \pi} +1)<\alpha < 0\), then \(g_\alpha(r)<\sqrt2\) and
\(4 \pi \alpha r\le -1\) for all \(r\ge a\).
Hence the condition $\omega(r, \alpha)  \leq \pi$ is satisfied.

2) If \(\alpha<-(\frac{1}{4 \sqrt{2} \pi} +1)\), we have trivial inequalities $-\frac{r}{\sqrt2}+e^{-r/\sqrt2}\cos\frac{r}{\sqrt2} \leq 1$ and $\frac{r}{\sqrt2}+e^{-r/\sqrt2}\sin\frac{r}{\sqrt2} \geq 0$ and see, $\omega(r, \alpha)  \leq 1, \, \alpha <-(\frac{1}{4 \sqrt{2} \pi} +1)$.

We conclude that for every \(\alpha < 0\) the map
\(r\mapsto u(r)=f^{-1}(\omega(r, \alpha))\) is analytic on the entire
half-line \([a,\infty)\), and its Taylor series at \(r=a\)
converges for all \(r\ge a\).

\section{Proof of Proposition~\ref{proposition_Efimov_constant_potential}}\label{appendix2}
By considering \eqref{scaling_energy}, we can rewrite the eigenvalue problem \eqref{reduced-radial-higher-momentum} as 
\begin{align}\label{eqb1}
& u^{\prime \prime}  -\left(\frac{l(l+1)}{r^2} + \eta ^2 + \ve^{-2}\Lambda \right) u =0 \;\;\;\;& \;\text{for} \;\; r<r_0  \\
& u^{\prime \prime} + \left( \frac{\beta ^2 + 1/4}{r^2} - \eta ^2\right)u = 0 \;\;\;& \text{for} \;\; r >r_0 \label{eqb2}
\end{align}
\noindent
with the boundary conditions 
\begin{equation}\label{bc}
    \begin{split}
        &\text{for} \,\,\, l=0, \quad  u(0)= 0 \\
        &\text{for} \,\,\, l \geq 1, \quad \lim_{r \to 0} u(r)= 0, \quad \lim_{r \to 0}r^{-(l+1)}u(r)=1
    \end{split}
\end{equation}
These equations are respectively modified Bessel equations of real order $l + 1/2$ and imaginary order $\nu=i\beta$. As a result, the general solution of \eqref{eqb1} is a linear combination of modified Bessel functions of the first and second kind. However, the modified Bessel function of the first kind $I$ with purely imaginary order becomes complex on the positive real axis, so it not admissible as the solution of \eqref{eqb2} that satisfies \eqref{bc}. Therefore, Dunster \cite{Du} introduces the modified Bessel function $L_\gamma$:
\begin{equation}\label{Bessel_L}
    L_\gamma(z)=\frac{\pi i}{2 \sin (\gamma \pi)}\left\{I_\gamma(z)+I_{-\gamma}(z)\right\} \quad(\gamma \neq 0)
\end{equation}
As a result, the general solution for \eqref{eqb2} can be written as a linear combination of the Bessel functions $L_{i\beta}(\eta r)$ and $K_{i \beta} (\eta r)$. Set $$\xi = (\sqrt{\eta^{2}+\ve^{-2}\Lambda}\, r),$$ we have the following form of the solutions:
    \begin{align}\label{so<}
&u (r<r_0) = A \,\sqrt{\xi } I_{l + 1/2} (\xi ) + B\, \sqrt{\xi } K_{l + 1/2} (\xi ) \\
&u(r>r_0) = C \, \sqrt{r}  L_{i\beta} (\eta\, r) + D\, \sqrt{r}  K_{i\beta}(\eta\, r), \label{so>}
\end{align}
where $A, B, C, D$ are arbitrary complex constants. Note that, the modified Bessel function $L_{i\beta}$ grows exponentially fast at infinity and the modified Bessel function $K_{l+1/2}$ is singular at the origin. Taking into account this property alongside the boundary conditions \eqref{bc}, we obtain $B=C=0$ in \eqref{so<} and \eqref{so>}. Thus,
\begin{equation}\label{solutions}
\begin{split}
    &u (r<r_0) = A\, \sqrt{\xi } I_{l + 1/2} (\xi ),\\
&u (r>r_0) = D\, \sqrt{r} K_{i\beta}(\eta r)\,.
\end{split}
\end{equation}
Let us denote $\tau_0= \eta r_0$ . The solution has to satisfy matching conditions, \emph{i.e.} the continuity of the function and its derivative in $r = r_{0}$. This condition implies that the coefficient $A, D$ have to satisfy:   
\begin{widetext}
    \begin{equation}\label{lisy}
\left\{ 
\begin{aligned} 
\sqrt{\xi} I_{l + 1/2} (\xi) \, &A & - &  \; \;\; \sqrt{r_0}K_{i\beta}(\tau_0) \,  D &=0\\
        \left( \frac{\sqrt{\xi}}{2r} I_{l + 1/2}(\xi) + \frac{\sqrt{\xi}\xi}{r} I_{l + 1/2}^{\prime}(\xi) \right) & A & - & \;\left( \frac{1}{2 \sqrt{r_0}} K_{i\beta}(\tau_0) + \eta \sqrt{r_0} K_{i\beta}^{\prime}(\tau_0) \right)  D &=0
\end{aligned}
\right.
\end{equation}
\end{widetext}
This linear homogeneous system has non-zero solutions if and only if its determinant is equal to zero. In other words, we have to satisfy: 
\begin{equation}\label{eqtr1}
    \begin{split}
        &K_{i\beta}(\tau_0)  =  2\tau_0 f(\xi)\, K_{i\beta}^{\prime}(\tau_0) \qquad \text{with} \\ 
        &f(\xi) :=\frac{I_{l + 1/2} (\xi)}{2\xi \, I_{l + 1/2}^{\prime} (\xi)} = \frac{I_{l + 1/2} (\xi)}{\xi \left(I_{l - 1/2} (\xi) + I_{l + 3/2} (\xi)\right)}\,.
    \end{split}
\end{equation}
\noindent
Note that by the asymptotic behaviors of the functions $I_{l \pm 1/2}$ (see \cite[section 9.6]{AS}) we have:
\begin{equation*}
    \;\lim_{\xi \rightarrow 0} f(\xi)=\frac{1}{2l+1}\, .
\end{equation*}
Furthermore,  for $\tau_0 \rightarrow 0$ we also have (see \cite{Du}):
\begin{equation} \label{behav_Bessel}
\begin{split}
    K_{i\beta}(\tau_0) &= C_{\beta} \sin \left( \beta \ln \frac{\tau_0}{2} - \phi_{\beta,0} \right) + O(\tau_0^2) \\
K_{i\beta}^{\prime}(\tau_0) &= C_{\beta} \frac{\beta}{\tau_0}  \cos \left( \beta \ln \frac{\tau_0}{2} - \phi_{\beta,0} \right) + O(\tau_0)
\end{split}
\end{equation}
where 
\begin{equation}
C_{\beta}= - \sqrt{\frac{\pi}{\beta \sinh (\beta \pi)}} \quad \text{and} \quad \phi_{\beta,0}= \arg \Gamma (1 + i \beta) \,.
\end{equation}
As a result, the equation \eqref{eqtr1} in the limit $\tau_0 \to 0$ (\emph{i.e.} $\eta \to 0$) can be rewritten as:
\begin{equation}\label{bi-H}
 \sin \left( \beta \ln \frac{\tau_0 }{2} - \phi_{\beta,0} \right) - 2 \beta f(\xi_0) \cos \left( \beta \ln \frac{\tau_0 }{2} - \phi_{\beta,0} \right) = F(\tau_0 )
\end{equation}
where $\xi_0$ was defined in \eqref{scaling_energy}, and $F$ is (at least) continuous and $F(\tau_0)=O(\tau_0^2)$ for $\; \tau_0 \rightarrow 0$. With the change of coordinates $\theta= \beta \ln \frac{\tau_0}{2} - \phi_{\beta,0}$ and the equation, we have \eqref{bi-H}:
\begin{equation}\label{bo-H}
    \theta = \arctan \left(2\beta \,f(\xi_0) + \frac{F\left( \tau_0 \right)}{cos \theta}\right)
\end{equation}
Note that $\frac{F(\tau_0)}{\cos{\theta}} $ with $\tau_0 = 2 \, \ee^{\tfrac{\phi_{\beta,0}}{\beta}}  \ee^{\tfrac{\theta}{\beta}}$ tends to zero with the order of $O\left(\tau_0^2\right)$ as $\tau_0 \rightarrow 0$. Also, $f(\xi)$, $\xi \in \mathbb{C}$ is analytic in the unit disk in vicinity of the origin.
\\
As a result, for any value of $r_{0}$, there exists an infinite sequence $\theta_n$ of solutions of equation \eqref{bo-H} as $\tau_0 \rightarrow 0$, with $\theta_n <0$ and $\displaystyle \lim_{n \to +\infty} \theta_n =- \infty$.
\\
To conclude the proof, we have to show that the sequence of $-(\eta_n)^2$, corresponds this sequence of $\theta_n$ as the solutions of \eqref{bo-H} can be written in the form \eqref{asei}. Let us denote by $-\left(\eta^0_n\right)^2$ as the solutions of \eqref{bo-H} corresponds to $F(\tau_0)=0$. We have:
\begin{equation}
- \left(\eta^0_n\right)^2 =\ve^{-2} E^0_n = - \frac{4}{ r_0^2} \, \ee^{ \, \frac{2}{\beta} \left(   \arctan \big(2\beta f\left(\xi_0\right)\big) + \phi_{\beta,0} -n \pi   \right)}
\end{equation}
It is evident that the function $\eta_n = \frac{2}{r_0}\,\ee^{\frac{\phi_{\beta,0}}{\beta}} e^{\frac{\theta_n}{\beta}}$ is a differentiable function of $\eta$ (see \eqref{bo-H}) with bounded derivative which converges to $0$ for $\eta_{n} \rightarrow 0$. Consequently, 
\begin{equation*}
    -(\eta_n)^2 = - (\eta^{0}_{n})^{2} + \zeta_n \, , 
\end{equation*}
where $\zeta_n \to 0$ as $n \to \infty$.
\section{Elements of Ulam stability}\label{Appendix_Ulam}
We follow \cite{ANDERSON2025128908} with some notational changes. Define a second-order differential operator $\mathcal{D}$ on $C^2(I,\mathbb{R})$ by
\begin{equation}\label{generic_Ulam_operator}
    \mathcal{D}(f) := a(r)f^{\prime \prime}(r)+ b(r)f^\prime(r)+c(r)f(r) - d(r) \, ,
\end{equation}
where $a,b,c,d$ are at least continuous on $I$. We say that $\mathcal{D}$ is Ulam stable on $I$ if there exists a constant $C$ such that, for every $\delta>0$ and every $h \in C^2(I,\mathbb{R})$ satisfying
\begin{equation}\label{Ulam_stability_condition}
    \sup_{r \in I} \big\lvert a(r)h^{\prime \prime}(r)+ b(r)h^\prime(r)+c(r)h(r) - d(r) \big\rvert \leq \delta \, ,
\end{equation}
there is a $C^2$ solution $f(r)$ of \eqref{generic_Ulam_operator} with 
\begin{equation}\label{Ulam_stability_def}
    \sup_{r \in I} \big| h(r) - f(r) \big| \leq C\delta \, .
\end{equation}
To prove \eqref{Ulam_stability_def}, one typically derives the Riccati equation associated with \eqref{generic_Ulam_operator}:
\begin{equation}\label{riccati}
    a(r)\left(\rho^\prime(r)+\rho^2(r)\right) + b(r)\rho(r) = 0\, .
\end{equation}
Let $I=(\gamma,\sigma)$ and define:
        \begin{equation}\label{func_kappa_1}
            \begin{split}
                \kappa_1 (r) :=& \int_r^\sigma \frac{\ee^{\int_r^\sigma \left(\rho(t) + \tfrac{b(t)}{a(t)}\right)\dd t}}{\lvert a(s)\rvert} \dd s, \\
                \kappa_2 (r) :=& \int_\gamma^r \ee^{\int_s^r \rho(t)\dd t} \dd s \, .
            \end{split}
        \end{equation}
If $\lVert\kappa_i \rVert_{\infty} < \infty, \, i=1,2$, then \eqref{generic_Ulam_operator} is Ulam stable on $I$ (see \cite[Theorem 3.1, (ii)]{ANDERSON2025128908}).
\begin{proof}[\textbf{Proof of Lemma~\ref{lemma_Ulam}}]
\mbox{}\\
 Technically, we cannot apply Ulam stability theory to the neighborhood of the origin $r=0$, where the associated Riccati equation is not well-defined ($\ln$ diverges). Therefore, we treat that interval separately. We scale the energy as $ -\ve^2\eta^2$. Let $V_k^{-1}(\eta^2)$ be the point where $V_k(r) + \eta^2 = 0$. We split $[0,r_k]$ into $[0,V_k^{-1}(\eta^2)]$ and $(V_k^{-1}(\eta^2), r_k]$ and analyze \eqref{reduced-radial-higher-momentum} on them separately.
    \begin{itemize}
        \item[-]\textbf{\textit{step 1}: \textit{the interval $\mathbf{[0,V_k^{-1}(\eta^2)]}$}:}
        \\
        Since $-\lambda_0(r)$ is analytic and bounded, $-\lambda_0(r) \in L^1 \cap L^{\infty}$ on the interval $[0,V_k^{-1}(\eta^2)]$. Therefore, we can write $f(r)$ as the solution of \eqref{reduced-radial-higher-momentum} with the effectivee potential $V_{\mathrm{eff}} = V_k$ in \eqref{k_potentials} on this interval as (see \emph{e.g.} \cite{10.1063/1.1703665}):
            \begin{equation}
                f(r) = f_0(r) + \big(f_0* \left(-\lambda_0 \cdot f \right)\big)(r) \, ,
            \end{equation}
        where the convolution $\big(f_0* \left(-\lambda_0 \cdot f \right)\big)(r)$ defined as 
            \begin{equation*}
                \int_0^{r} -\lambda_0(t)f(t) \, f_0(r-t)  \dd t, \quad t>0 \, .
            \end{equation*}
        From the proof of Proposition \ref{proposition_Efimov_constant_potential}, $f_0(r) = A \, \sqrt{\eta  r} I_{l + 1/2} (\eta r)$, with $I_{l + 1/2}$ the modified Bessel function of the first kind and $A \in \cc$. Set $\tau = \eta r$. As $\tau \to 0$,
            \begin{equation}\label{BesselI_asym}
                \sqrt{\tau} I_{l + 1/2} (\tau) = \frac{1}{2l+1}\sqrt{\frac{2}{\pi}} \, \tau^{l+1} + O( \tau^{l+3}) \, .
            \end{equation}
        Without loss of generality take $A \in \rr^+$. Then, by \eqref{eqb1}, $f_0, f_0^\prime,f_0^{\prime\prime} > 0$ on $[0,V_k^{-1}(\eta^2)]$. On this interval $\eta^2 + V_k^{-1}(\eta^2) \geq 0$ by definition, so similarly $f, f^\prime,f^{\prime\prime} > 0$. Hence $\big(f_0* \left(-\lambda_0 \cdot f \right)\big) < 0$, which implies $|f_0(r)| \geq |f(r)|$ on $[0,V_k^{-1}(\eta^2)]$. Moreover, for $\tau \to 0$, \eqref{BesselI_asym} yields
           
                \begin{equation}\label{convolution_ineq}
                    \int_0^{V_k^{-1}(\eta^2)} -\lambda_0(t)f(t) \, f_0(r-t)  \dd t \leq L_1 \, \tau^{l+1} \, ,
                \end{equation}
        where the constant
                \begin{equation*}
                    L_1 = V_k^{-1}(\eta^2) \displaystyle \sup_{r \in [0,V_k^{-1}(\eta^2)]} \big|\lambda_0(r)f(r)\big| \, .
                \end{equation*}
        \item[-]\textbf{\textit{step 2}: \textit{the interval $\mathbf{(V_k^{-1}(\eta^2), r_k]}$}:}
            \\
        Here we apply Ulam stability directly. Substituting \eqref{reduced-radial-higher-momentum} into \eqref{generic_Ulam_operator} with $ E = -\ve^2\eta^2$ gives
                \begin{equation*}
                   \begin{split}
                        &a(r) = 1, \;\;\; b(r) = 0, \;\;\; d(r) =0 \\ &c(r) = \ve^{-2}\lambda_0(r) - \frac{l(l+1)}{r^2} - \eta^2 \, .
                   \end{split}
                \end{equation*}
       Since we have excluded $r=0$, $c(r)$ is not singular. From Lemma~\ref{analyticity_lemma}, $-\lambda_0(r)$ is analytic and the radius of convergence of its Taylor series about the origin is infinite. This fact, implies that by the Cauchy–Kovalevskaya theorem (see, \emph{e.g.}, \cite{Gaidashev_CK}), the solution $f(r)$ is unique and analytic on $(V_k^{-1}(\eta^2), r_k]$. Consequently, the functions $\kappa_i, \, i=1,2$ in \eqref{func_kappa_1} are bounded, and \eqref{reduced-radial-higher-momentum} with $V_{\mathrm{eff}} = V_k$ is Ulam stable on this interval. Moreover, $f_0(r) = A \, \sqrt{\eta  r} I_{l + 1/2} (\eta r)$ is bounded here, and for low energies $\eta \to 0$ there exists a constant $L_2$ such that \eqref{Ulam_stability_condition} holds:
            \begin{equation*}
               \begin{split}
                    &\Big| f_0^{\prime\prime}(r) + \left(\ve^{-2}\lambda_0(r) - \frac{l(l+1)}{r^2} - \eta^2\right)f_0(r) \Big| =\\  &\Big| \ve^{-2}\lambda_0(r) f_0(r) \Big| \leq L_2 (\eta r)^{l+1} \, .
               \end{split}
            \end{equation*}
        Therefore, there exists a constant $L_3$ on $(V_k^{-1}(\eta^2), r_k]$ such that
            \begin{equation*}
                \big\lvert f(r) - f_0(r) \big\rvert \leq L_3 \left(\eta r\right)^{l+1} \, .
            \end{equation*}
    \end{itemize}
    Combining the two steps and setting $C_k := max\{L_1, L_3\}$ completes the proof.
\end{proof}
Note that, in view of the asymptotic behavior of $-\lambda_0(r)$ at $r=0$ displayed in \eqref{E01}, it may be possible to improve the exponent $l+1$ in \eqref{convolution_ineq}. Moreover, the second step of the proof of Lemma~\ref{lemma_Ulam} can still be carried out under weaker regularity assumptions on $-\lambda_0(r)$.
\section{Proof of Proposition~\ref{proposition_non_Efimov_eig}}\label{App_non_Efimov}
    Let $u^l_{c}(r)$ denote the solution of \eqref{reduced-radial-higher-momentum} with the potential \eqref{auxiliary_potential} and $\Lambda = -\lambda_0(r_{\mathrm{min}})$. Then, from \eqref{solutions}, using standard Bessel identities (see \cite[section 9]{AS}) and the scaling \eqref{scaling_energy}, 
   \begin{widetext}
        \begin{equation}
        u^l_{c}(r) = A \left(\sqrt{\ve^{-2}\,\lambda_0(r_{\mathrm{min}}) -\eta^2}\, r\right)^{1/2} J_{l + 1/2} \left(\sqrt{\ve^{-2}\,\lambda_0(r_{\mathrm{min}}) -\eta^2}\, r\right) \,, \;\;\;\;\;\;\;\; r<r_0 \,.
    \end{equation}
   \end{widetext}
    Because $|\lambda_0(r)| \leq |\lambda_0(r_{\mathrm{min}})|$, the Sturm–Picone comparison theorem \cite{diaz1969sturm} on $[0,r_0]$ implies that the number of zeros of $u^l_c$ is greater than that of $u$ in \eqref{solutions}. Hence, for $l \geq 0$, if the first zero of $u^l_c(r)$ (besides $u_c(0)=0$), occurring at $r_c$, lies outside $[0,r_0]$, there is no eigenvalue on this interval. Equivalently,
    \begin{equation}
        \sqrt{\ve^{-2}\,\lambda_0(r_{\mathrm{min}}) - \eta^2} \, r_c \leq  \mathscr{j}_{l+1/2,1} \, .
    \end{equation}
    Setting $\eta = 0$ and $r_c = \sqrt{2} \frac{3 \pi}{4}$ yields \eqref{non-Efimov_mass_ratio_condition}.
\section{Proof of Lemma~\ref{Lemma_second_order_error}}\label{App_Lemma_second_order}
    Similar to the proof of Proposition~\ref{proposition_Efimov_constant_potential}, we have to solve the equation (see also \eqref{eqtr1})
 \begin{equation}\label{second_order_eqtr1}
     K_{i\beta}(\tau_0)  =  \frac{2\tau_0}{2l+1} \, K_{i\beta}^{\prime}(\tau_0)
 \end{equation}
    Recall the definition of the modified Bessel function of the second kind, $K_{i \beta}(x)$ (see \emph{e.g.}\cite[Eq. (2.8)]{Du})
        \begin{equation} 
            -\left(\frac{\beta \pi}{\sinh (\beta \pi)}\right)^{1 / 2} \sum_{s=0}^{\infty} \frac{\left(x^2 / 4\right)^s \sin \left(\beta \ln (x / 2)-\phi_{\beta, s}\right)}{s!\left[\left(\beta^2\right)\left(1^2+\beta^2\right) \cdots\left(s^2+\beta^2\right)\right]^{1 / 2}} \, ,
        \end{equation} 
    where 
        \begin{equation}
            \phi_{\beta,s}= \arg \Gamma (1 + s + i \beta) \,.
        \end{equation}
    Denoting $C_{\beta}= - \left(\frac{\pi}{\beta \sinh (\beta \pi)}\right)^{1/2}$, we obtain the asymptotic behavior \eqref{behav_Bessel} for for $\tau_0 \to 0$ in the leading order. To calculate the next order, by using the identity
 \begin{equation}
     \phi_{\beta,1} = \phi_{\beta,0} + \arctan \beta \,
 \end{equation}
 we have
 \begin{widetext}
     \begin{equation} 
    \begin{split}
        K_{i\beta}(\tau_0) &= C_{\beta} \left[\left(1+ \frac{\tau_0^2}{4(1+\beta^2)}\right)\sin \left( \beta \ln \frac{\tau_0}{2} - \phi_{\beta,0} \right) \, -\frac{\beta\tau_0^2}{4(1+\beta^2)}\cos \left( \beta \ln \frac{\tau_0}{2} - \phi_{\beta,1} \right)\right] + O(\tau_0^4) \\
    K_{i\beta}^{\prime}(\tau_0)&= C_{\beta} \left[\frac{\beta}{\tau_0} \cos{\left( \beta \ln \frac{\tau_0}{2} - \phi_{\beta,0} \right)} + \frac{\tau_0}{4(1+\beta^2)} \left((2 + \beta^2)\sin \left( \beta \ln \frac{\tau_0}{2} - \phi_{\beta,0} \right) - \beta \cos{\left( \beta \ln \frac{\tau_0}{2} - \phi_{\beta,0} \right)}\right) \right] + O(\tau_0^3)
    \end{split}
\end{equation}
 \end{widetext}
 Setting $\theta_0 = \beta \ln \frac{\tau_0}{2} - \phi_{\beta,0}$, from \eqref{second_order_eqtr1} we have
 \begin{equation*}
     \tan \theta_0 = \frac{2 \beta}{2l+1} + O(\tau_0^3) \, .
 \end{equation*}
The solution has a form $\theta = \theta_0 + \delta$. The rest of the proof is a straightforward linearization to find the exact from of $\delta$. We define
\begin{equation*}
    f(\theta) = \sin \theta - \frac{2\beta}{2l+1} \cos \theta \, .
\end{equation*}
By construction $f(\theta_0) = O(\tau_0^3)$. Moreover, $f$ is smooth and $f(\theta_0+ \delta) = f^{\prime}(\theta_0)\delta + O(\delta^2) $. Therefore, we can re-write \eqref{second_order_eqtr1} 
\begin{widetext}
    \begin{equation*}
    f^{\prime}(\theta_0)\delta + \tau_0^2 \left(\frac{2l-3-2\beta^2}{4(2l+1)(\beta^2+1)} \sin \theta_0 - \frac{\beta (2l-1)}{4(2l+1)(\beta^2+1)} \cos \theta_0\right) = O(\tau_0^3) \, .
\end{equation*}
\end{widetext}
By the same argument as the proof of Proposition~\ref{proposition_Efimov_constant_potential}, we obtain \eqref{second_order_delta} and conclude the proof.
\section*{Acknowledgements}
I am especially grateful to Rodolfo Figari for his guidance and many helpful discussions that contributed to the development of this work. I also thank Andrea Posilicano, Claudio Cacciapuoti, and Alessandro Teta for constructive conversations that supported various parts of this work.

The Author also acknowledge funding from the Next Generation EU-Prin project 2022CHELC7 “Singular
Interactions and Effective Models in Mathematical Physics” and the support of the National Group of
Mathematical Physics (GNFM-INdAM).

\bibliography{references}

\end{document}